\documentclass{aa} 

\usepackage{natbib}
\usepackage{graphicx}
\usepackage{caption}
\usepackage{subcaption}
\usepackage{rotating}
\usepackage{color}
\usepackage[varg]{txfonts}
\usepackage{amssymb,amsmath}
\usepackage{textcomp}
\usepackage{epstopdf}
\usepackage{float}
\usepackage{hyperref}
\hypersetup{colorlinks,%
citecolor=blue,%
filecolor=blue,%
linkcolor=blue,%
urlcolor=blue}
\usepackage{url}

\begin{document} 
\title{Variability of young stellar objects in the star-forming region Pelican Nebula }

\author{A. Bhardwaj\inst{1}
          \and 
          N. Panwar\inst{2}
          \and
          G. J. Herczeg\inst{1}
          \and
          W. P. Chen\inst{3}          
          \and 
          H. P. Singh\inst{4}
           }
\institute{Kavli Institute for Astronomy and Astrophysics, Peking University, Yi He Yuan Lu 5, Hai Dian District, Beijing 100871, China\\
   	   \email{anupam.bhardwajj@gmail.com; abhardwaj@pku.edu.cn}
         \and
		Aryabhatta Research Institute of Observational Sciences, Manora Peak, Nainital-263002, Uttarakhand, India
	 \and
	     Graduate Institute of Astronomy, National Central University, Jhongli 32001, Taiwan
         \and
             Department of Physics and Astrophysics, University of Delhi,  Delhi-110007, India
             }
\date{Received xxxx; accepted xxxx}

  \abstract
   {Pre-main-sequence variability characteristics are useful to probe the physical processes leading to the formation and initial evolution of both stars and planets.}
{Photometric variability of pre-main-sequence stars is studied at optical wavelengths to explore star-disk interactions, accretion, spots and other physical mechanisms 
associated with young stellar objects.
}
{We observed a field of $16'\times 16'$ in the star-forming region Pelican Nebula (IC 5070) at $BVRI$ wavelengths for 90 nights spread over one year in 2012-2013. More than
250 epochs in $VRI$-bands are used to identify and classify variables up to $V\sim 21$~mag. Their physical association with the cluster IC 5070 is established based on the parallaxes and proper 
motions from the {\it Gaia} second data release. Multiwavelength photometric data are used to estimate physical parameters based on the isochrone fitting and spectral energy distributions.}
{We present a catalogue of optical time-series photometry with periods, mean-magnitudes and classifications for 95 variable stars including 67 pre-main-sequence variables
towards star-forming region IC 5070. The pre-main-sequence variables are further classified as candidate classical T Tauri and weak-line T Tauri stars
based on their light curve variations and the locations on the color-color and color-magnitude diagrams using optical and infrared data together with Gaia DR2 astrometry.
Classical T Tauri stars display variability amplitudes up to three times the maximum fluctuation in disk-free weak-line T Tauri stars, which show strong periodic
variations. Short-term variability is missed in our photometry within single nights. Several classical T Tauri stars display long-lasting ($\geq 10$ days) single or multiple fading 
and brightening events up to a couple of magnitudes at optical wavelengths. The typical mass and age of the pre-main-sequence variables from the isochrone-fitting and spectral energy
distributions are estimated to be $\le 1~M_\odot$ and $\sim 2$ Myr, respectively. We do not find any correlation between the optical amplitudes or periods with the physical 
parameters (mass and age) of pre-main-sequence stars.}
{The low-mass pre-main-sequence stars in the Pelican Nebula region display distinct variability and color trends and nearly $30\%$ variables exhibit strong
periodic signatures attributed to cold spot modulations. In case of accretion bursts and extinction events, the average amplitudes are larger than one magnitude 
at optical wavelengths. These optical magnitude fluctuations are stable over the year-long time-scale.} 


\titlerunning{Variable stars in IC 5070}
\maketitle
%

\section{Introduction}
Most stars show variability in their brightness during some stage of their life-cycle. Photometric variability is a characteristic feature of the stars 
in their pre-main-sequence (PMS) phase and provides an insight into the different physical processes in young stars when studied at multiple wavelengths.
Variability is a ubiquitous property of T-Tauri stars (TTSs) that are low-mass ($M<2M_\odot$) PMS objects \citep{joy1945}.
``Classical'' TTSs (CTTSs) actively accrete material from the circumstellar disks while ``Weak-line'' TTSs (WTTSs) do not show any on-going 
accretion perhaps due to the lack of the inner disks. CTTSs exhibit large photometric variability with excess infrared and ultraviolet emission, 
and strong $H_\alpha$ emission. In contrast, WTTSs show periodic variability with smaller amplitudes, little or no infrared excess and a smaller $H_\alpha$ 
equivalent width \citep{herbig1962, herbig1977, bertout1989, herbst1994}.   Herbig Ae/Be represent more massive ($2M_\odot < M < 8M_\odot$) class of PMS stars 
and exhibit different types of photometric variability as they evolve towards the zero-age main-sequence (ZAMS). Some of the massive stars that reach 
the main-sequence (MS) in their core hydrogen burning phase, also display changes in their brightness due to pulsations, for example, $\beta$ Cep, $\delta$ Scuti 
or slowly pulsating B stars \citep{eyer2008}.  

Optical photometric variability of young stellar objects (YSOs) is attributed to a range of physical mechanisms.
Variability in WTTSs occurs due to an asymmetric distribution of cool or dark magnetic spots at the stellar surface 
that modulates the observed luminosity of the star during its rotation \citep{bouvier1993, herbst1994, grankin2008}. 
In CTTSs, variability is caused by the variable accretion from the circumstellar disk onto the star, where both the accretion rate and the 
distribution of accretion zone or hot-spots over the stellar surface are not uniform \citep{herbst2007, cody2014}. 
Variability in Herbig Ae/Be stars predominantly occurs due to the obscuration from the circumstellar dust \citep{bertout1989, herbst1994, 
herbst2007, semkov2011, stelzer2015}. Since, YSOs exhibit different variability signatures, exploring their variable properties at multiple wavelengths  
is essential to understand the complex nature of these stars in both short and long time-scales. 

Several studies have focussed on the PMS variability of YSOs with an aim to understand the star-disk interactions, accretion, outflows and other physical mechanisms
\citep[e.g.,][]{grankin2008, alencar2010, venuti2015, messina2017, rodriguez2017, fernandes2018, guo2018}. Space-based observations allowed a remarkable progress in YSO variability studies thanks to
the high-precision photometry that probes the flux variation to 1$\%$ of amplitudes and over sub-hour time-scales \citep{alencar2010, cody2014, ansdell2016, gillen2017}. 
Using photometric data with unprecedented accuracy, a detailed (sub)-classification of YSOs into quasi-periodic, dippers, bursters etc. was provided by \citet{cody2014} based on 
their light curve morphology at multiple wavelengths. \citet{venuti2015} studied the variability and accretion dynamics of YSOs in the NGC 2264 at
ultraviolet and optical wavelengths and found that accretion process is stable over long time-scales of years. The PMS variability could also contribute to the large 
scatter observed in Hertzsprung-Russell diagrams for star-forming regions \citep[SFRs,][]{baraffe2009, baraffe2012}, while the correlation of the rotation period/amplitude 
with different stellar properties can potentially provide insight into the angular momentum evolution in PMS stars \citep{bouvier1997,herbst2007}. 

The North American (NGC 7000) and Pelican (IC 5070) Nebulae are SFRs that are within one kiloparsec of the Sun. This  
complex provides an ideal laboratory to study the influence of massive stars on subsequent star formation activity and evolution of the
natal molecular clouds \citep{gui2009, rebull2011, zhang2014, bally2014}. These regions possess a large number of young PMS stars,
cometary nebulae, bright-rimmed clouds, collimated jets and Herbig-Haro objects \citep{ogura2002, ikeda2008, rebull2011, panwar2014, bally2014}. 
\citet{rebull2011} identified more than 2000 YSOs in the 7 deg$^2$ field towards North American and Pelican complex including
nearly 250 YSOs in the Pelican cluster. However, the long-term optical photometric studies of PMS stars  
in these regions are available only for a limited sample \citep[][and references within]{kospal2011, findeisen2013, poljan2014, Ibryamov2018}. 
The $BVRI$ photometry for a sample of 17 PMS objects was presented by 
\citet{poljan2014} in the field of North American and Pelican Nebulae while \citet{froebrich2018} found two new low-mass young stars with deep recurring eclipses in IC 5070. 

In this work, we present a relatively large sample of variable YSOs in the IC 5070 based on a year-long optical photometry. 
The paper is structured as follow: Section~\ref{sec:obs} provides details of the observations, data reduction, photometric and astrometric calibrations. The variability identification, 
period determination and a comparison with published work are discussed in Section~\ref{sec:var}. The classification of YSOs based on their kinematics, color-color and color-magnitude
diagrams, and the light curve variations are discussed in Section~\ref{sec:class}. A detailed discussion on the physical and variable characteristics of PMS stars is presented in
Section~\ref{sec:pms} including the estimates of physical parameters based on the spectral energy distributions fitting tool. Final results of this work are summarized in 
Section~\ref{sec:summary}.

\section{Observations, data reduction, photometric and astrometric calibrations}
\label{sec:obs}

\begin{figure}
\centering
\includegraphics[width=1.0\columnwidth]{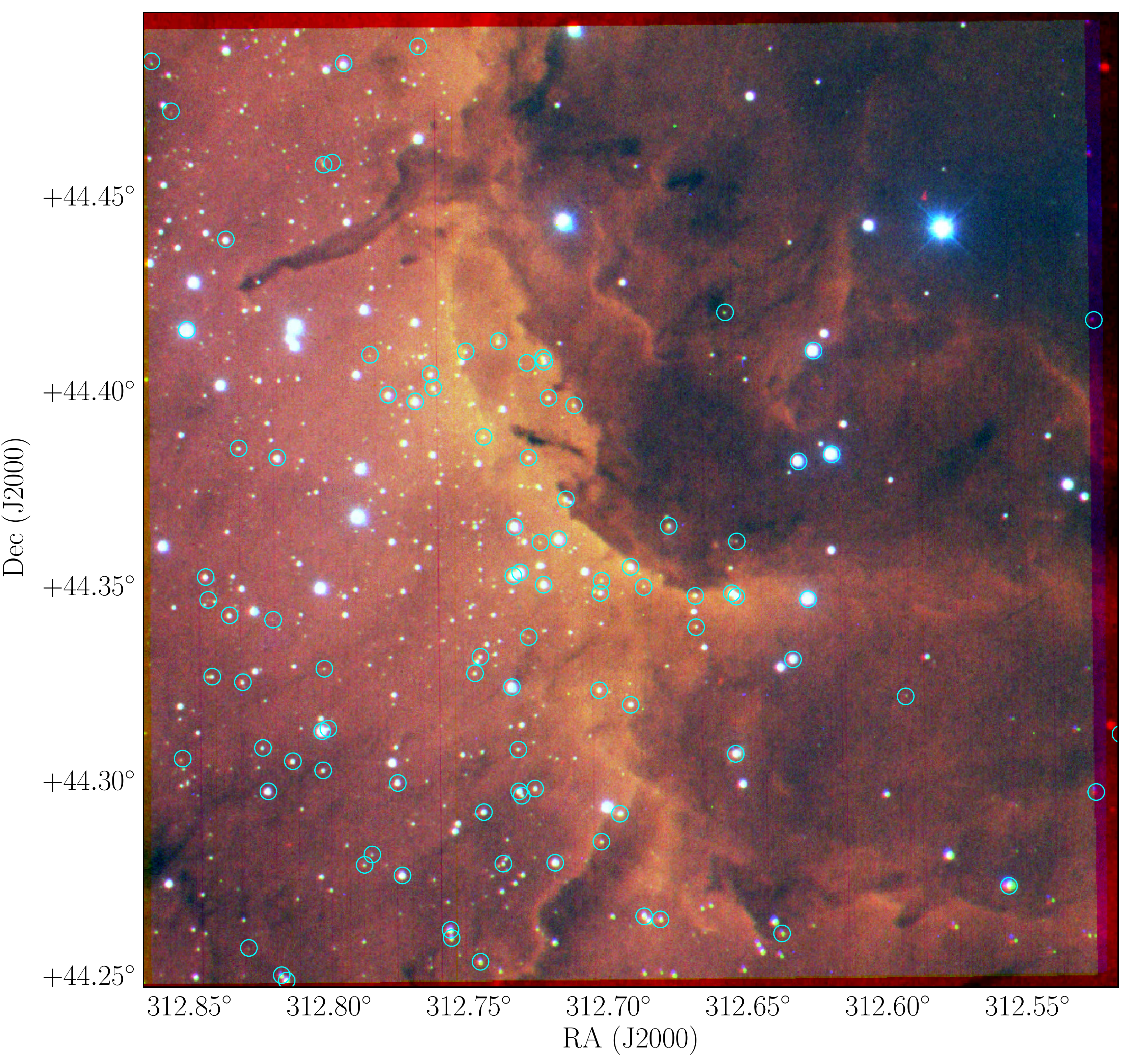}
\caption{Color composite image of the young star-forming region towards Pelican Nebula obtained using the $VRI$-band images. The cyan circles represent the location of
the variable stars.}
\label{fig:rgb}
\end{figure}

The observations were carried out using the 0.81-m ($32''$) Tenagra telescope that uses a science camera with 2048$\times$2048 pixels having 
an effective plate scale of $\sim 0.98''$ per pixel and a field of view of 16.8$\times$16.8 arc-minutes$^2$. The images in $VRI$ filters were acquired between 
May 2012 to June 2013 for 90 nights, often thrice each night but within one hour, while the $B$-band images were taken only on two nights. There are only 5 frames in $B$ and 
around 250 frames in $VRI$, taken with exposures varying from 420s in $B$ to 90s in $I$.  Calibration images (bias and flats) were obtained 
nightly and the pre-processing of images (bias subtraction, flat-fielding etc.) was done in IRAF{\footnote{\url{http://iraf.noao.edu/}}}. Finally, 
760 scientifically useful images are used to perform photometry. Fig.~\ref{fig:rgb} shows the color-composite image of the selected star-forming region 
towards IC~5070 obtained using the $VRI$ images taken on the first night.

The time-series photometry of the processed images is performed using DAOPHOT/ALLSTAR \citep{stetson1987} and DAOMATCH/DAOMASTER \citep{stetson1993} routines.
All sources above the $4\sigma$ threshold in each image are selected and the aperture photometry is obtained with a radius of 4 pixels. Point-spread function (PSF)
is determined from 20 bright and isolated stars in each image that is used to perform the PSF photometry using ALLSTAR on all images.
The frames taken on the first-night are selected as the master image in each filter and the PSF photometry are used as input to DAOMATCH to derive accurate frame-to-frame coordinate 
transformations. These transformations are used to obtain the corrected magnitudes of the stars relative to their magnitudes in the master image for all frames using DAOMASTER. 
The photometry includes 1307 sources with more than 50 observations in the target field, and over 1000 stars with both $V$- and $I$-band measurements. Within our target 
region, optical counterparts of 77 of the $\sim 135$ YSOs from \citet{rebull2011} are found within $3''$ matching radius in our photometry.

\begin{figure}
\centering
\includegraphics[width=0.5\textwidth]{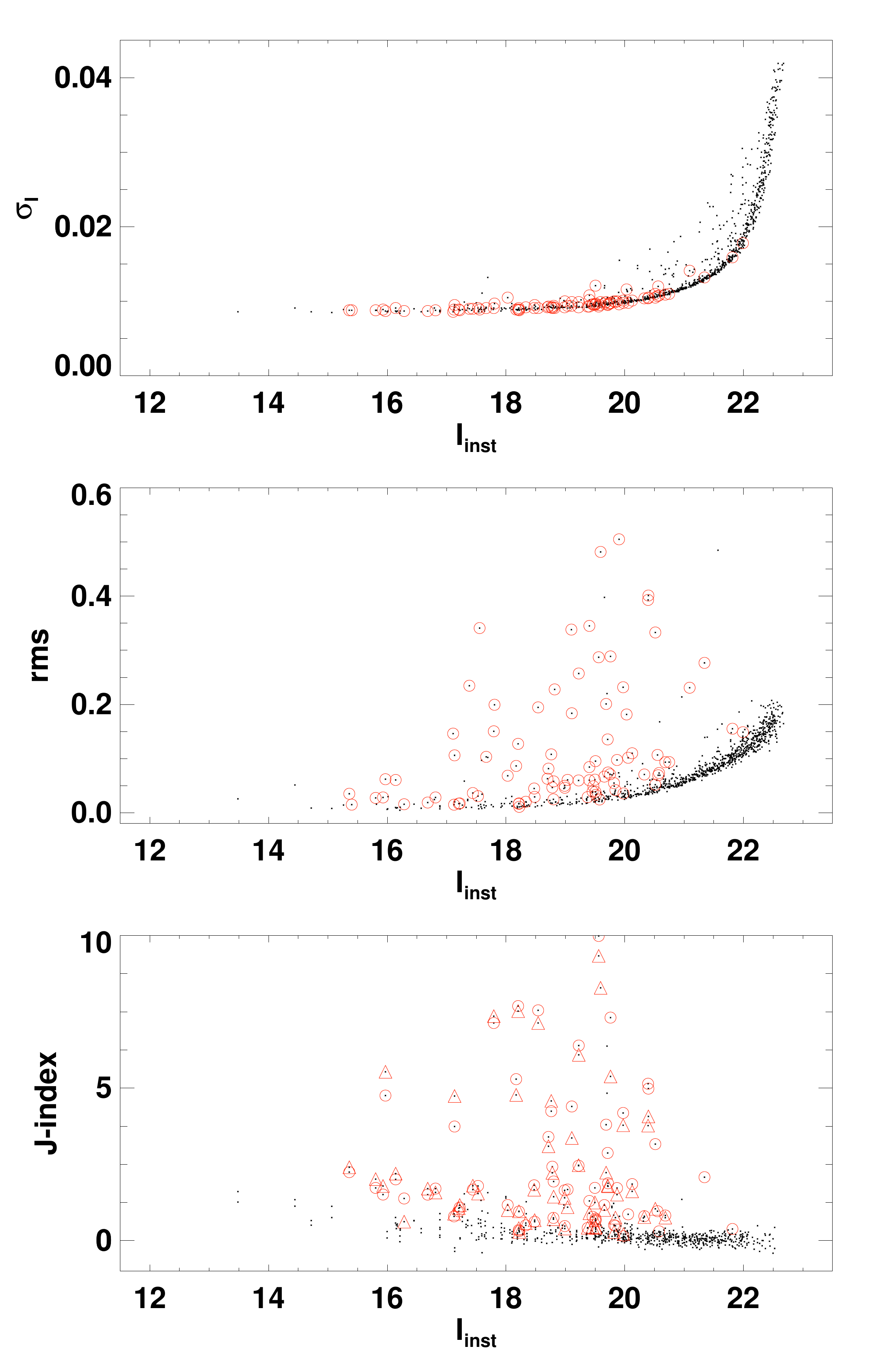}
\caption{{\it Top panel:} Photometric precision of our observations in $I$-band as a function of instrumental magnitudes. {\it Middle panel:} Root-mean-square (rms) 
scatter in $I$-band as a function of instrumental magnitudes. Red circles denote the selected candidate variables in the top and middle panels.
{\it Bottom panel:} Stetson's $J$-index for all sources in our field as a function of instrumental $I$-band magnitudes. Red circles/triangles represent the variables with 
light curves available in $RI/VI$ bands.}
\label{fig:inst_sig}
\end{figure}

The photometric calibration to Landolt filters \citep{landolt2009} are performed using the standard star photometry carried out on the same night as of the master frame.  
The standard transformation equations are:
\begin{eqnarray*}
b-B &=& 3.251   + 0.253 X_b - 0.056  (B-V),\\ \nonumber
v-V &=&  3.076  + 0.132 X_v + 0.018  (B-V),\\ \nonumber
r-R &=& 2.760   + 0.109  X_r + 0.127  (R-I),  \\ \nonumber
i-I &=& 3.863   + 0.033  X_i -0.007 (R-I), \nonumber
\label{eq:phot_cal}
\end{eqnarray*}

\noindent where, $b,v,r,i$ are instrumental magnitudes, $B,V,R,I$ are standard magnitudes and $X_b$,$X_v$,$X_r$,$X_i$ are the airmass in $BVRI$ filters, 
respectively. The maximum uncertainty of these coefficients and the dispersion in these relations are of the order of ~0.02 magnitude. If the
star is observed only in one filter, median instrumental color at the corresponding magnitude, is adopted to calibrate its magnitude. The top panel of Fig.~\ref{fig:inst_sig} shows the 
photometric precision of our observations as a function of $I$-band instrumental magnitude. For the calibrated magnitudes, the minimum uncertainty is 
0.02~mag for the brightest sources and exceeds 0.05 mag for the fainter targets. The calibrated magnitudes are compared
with a sample of 37 common stars with AAVSO Photometric All Sky Survey catalog \citep{henden2016} that resulted in a median absolute deviation of $\sim 0.2$~mag in $B$ and 
$\sim 0.1$~mag in $V$. We also compare the $BVRI$ magnitudes with small sample of stars compiled by \citet{gui2009} and \citet{findeisen2013}, and the 
median offset in each filter is found to be of the order of $10\%$ of the magnitude-range. The typical uncertainties in all the literature magnitudes
are also of the order of one tenth of a magnitude. The astrometric calibration is performed with ten bright and isolated sources using the Gaia DR2 data \citep{gaia2018}.

\begin{figure*}
\centering
\includegraphics[width=1.0\textwidth]{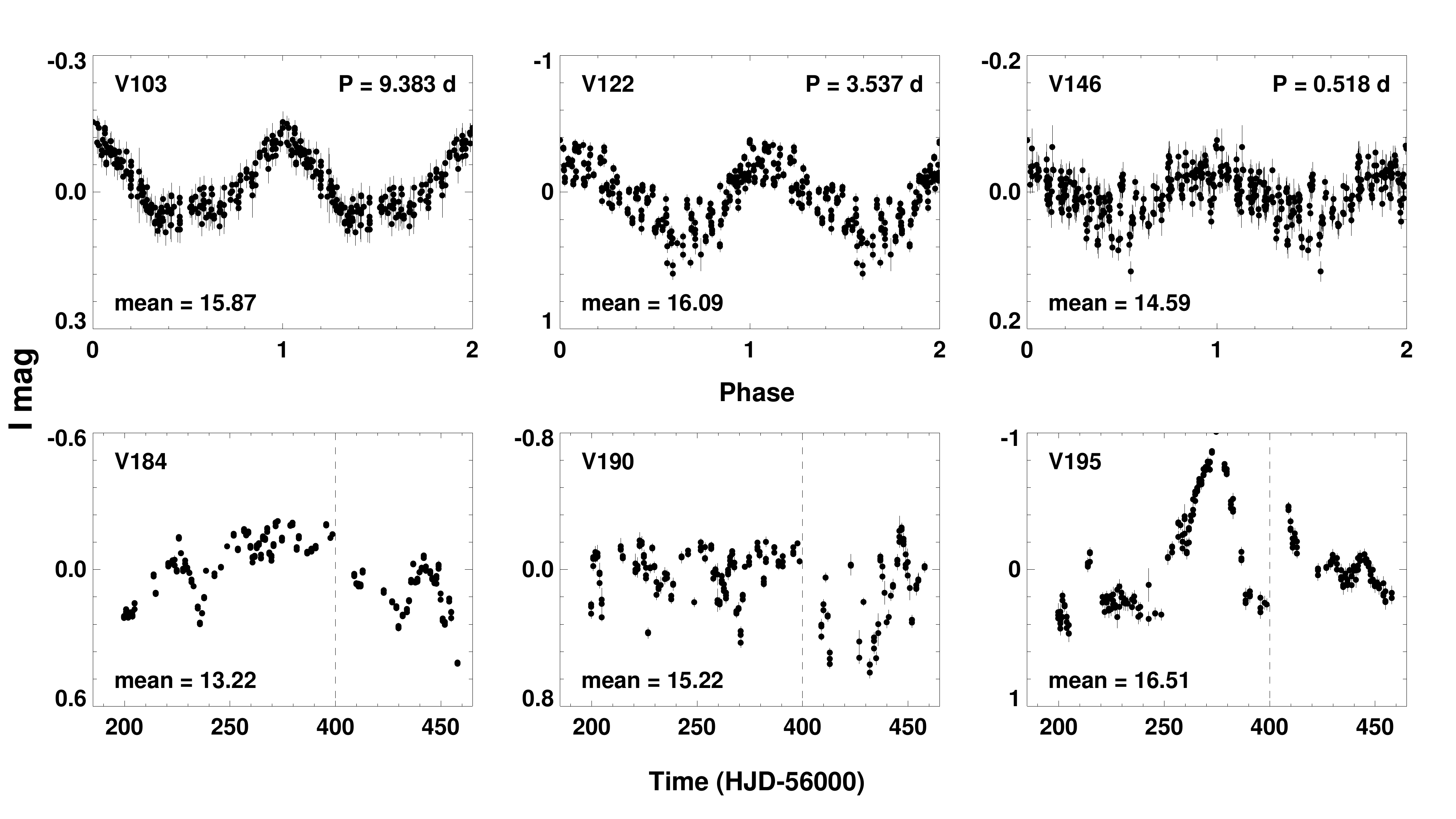}
\caption{The representative $I$-band light curves of periodic (top) and non-periodic variables (bottom) with varying light curve quality. The magnitudes are normalized with respect
to the zero-mean. Star ID, period and mean-magnitude are also provided in each panel. The vertical dashed line in the bottom panels 
separates the observations in two different seasons that are offset for visualization purposes.}
\label{fig:lcs}
\end{figure*}

The five-parameter solution (astrometry, parallax and proper motions) for all sources are obtained using a $3''$ search radius from Gaia DR2 catalog \citep{gaia2018cat}. 
Our astrometric calibration is not expected to be better than 1-2$''$ and therefore, all pairs within $1.5''$ are investigated to remove possible duplicates after combining 
data in different filters. Finally, the RA and DEC of the nearest neighbour source in the Gaia catalogue are adopted for further analysis. We also obtain various physical 
parameters including luminosity and effective temperatures, if provided in the Gaia catalogue. Note that \citet{luri2018} suggested
using a Bayesian approach to properly account for the covariance uncertainties in the parallaxes and proper motions from Gaia DR2. Therefore, accurate distances for these sources 
are also adopted from the catalog of \citet{coryn2018} that are determined using a Bayesian inference method based on a distance prior that varies smoothly as a function of Galactic 
longitude and latitude according to a Galaxy model.
Multiband photometric data for all the sources are also obtained by 
a cross-match within a search radius of $3''$ to the Two Micron All Sky Survey \citep[2MASS,][]{cutri2003}, {\it Spitzer}, Multiband Imaging Photometer for Spitzer (MIPS), and 
Wide-field Infrared Survey Explorer (WISE) archival catalogues{\footnote{\url{http://irsa.ipac.caltech.edu/applications/Gator/}}}.

\section{Variability search and period determination}
\label{sec:var}

We consider more than 1300 stars that are observed in at least 50 frames for the variability classification and period determination. At first,
the stars with large root-mean-square scatter around the mean magnitude from the combined multi-frame data are identified as candidate variables
in each filter. The dispersion around mean-magnitude from the time-series data provides a measure of the intrinsic variability of a 
star. For variable objects, root-mean-square scatter is significantly larger than the photometric noise while
the fluctuations around the mean are of the order of photometric uncertainties for the non-variable stars. Therefore, the full magnitude range 
are binned in different stepsizes of 0.2/0.5/1.0 mag and all stars above 3$\sigma$ level in each bin are selected as candidate variables.
Only stars for which the root-mean-square dispersion in $I$-band exceeds 0.05 mag are considered as variable candidates.
After combining individual variables in $VRI$-filters, a sample of 152 candidate variables show variability in at least one filter. 
Further, the correlated variability in different wavelengths is studied with a more robust approach using $J$-index \citep{stetson1996}. 
Stetson's $J$-index is calculated for stars with light curves available in the $VI$ and $RI$ filters using following equations:

\begin{eqnarray*}
P_k &=& \frac{N}{N-1} \left(\frac{I_k-\langle I \rangle}{\sigma_k(I)}\right) \left(\frac{V_k-\langle V \rangle}{\sigma_k(V)}\right),  \\ \nonumber
J   &=& \frac{\sum_{k=1}^{N} w_k~\text{sgn}(P_k) \sqrt |P_k|}{\sum_{k=1}^{N} w_k}, \nonumber
\label{eq:jindex}
\end{eqnarray*}

\noindent where, $P_k$ is the product of normalized residuals of $N$ pairs of simultaneous observations and {\it sgn} is the sign function. The weight $w_k$ is adopted as the inverse of the 
time difference between the pair of observations. The middle and bottom panels of Fig.~\ref{fig:inst_sig} show the rms scatter and the Stetson's $J$-index as a
function of instrumental $I$-band magnitudes. Most candidate variables selected from the rms scatter also have a $J$-index $\ge 0.5$ and no additional variable
candidate is found based on $J$-index value above this threshold.

\begin{table}
\caption{Time-series $VRI$-band photometry of variable sources.}
\begin{center}
\scalebox{1.0}{
\begin{tabular}{c c r r r}
\hline
ID & Band & MJD & Mag. & Error \\
\hline
V101& V& 56075.949&     15.373&      0.008\\
V101& V& 56075.952&     15.382&      0.008\\
V101& V& 56075.955&     15.377&     0.007\\
--- & ---&	---&	---&	---\\
V101& R& 56075.957&     14.830&      0.012\\
V101& R& 56075.958&     14.825&      0.010\\
V101& R& 56075.960&     14.819&      0.011\\
--- & ---& 	---&	---&	---	\\
V101& I& 56075.962&     14.323&      0.011\\
V101& I& 56075.963&     14.330&      0.009\\
V101& I& 56075.965&     14.327&      0.011\\
--- & ---&	---&	---&	---	\\
\hline
\end{tabular}}
\end{center}
{\footnotesize \textbf{Notes:} This table is available entirely in a machine-readable form in the online journal as supporting information. Only first-three
lines in each band are shown here for guidance regarding its form and content.}
\label{tab:phot}
\end{table}

\begin{table*}
\caption{Properties of variable sources.}
\begin{center}
\scalebox{0.85}{
\begin{tabular}{c c r r r r r r r r r r r r r}
\hline
ID & Ra & Dec & Period & \multicolumn{4}{c}{Mean-magnitudes} & \multicolumn{4}{c}{$\sigma$}& \multicolumn{3}{c}{Amplitudes} \\ 
   & deg. & deg. & days & B & V & R &I & B & V & R &I & $\Delta V$ & $\Delta R$ & $\Delta I$\\
\hline
V101&  312.77908&   44.39896&     22.238&     16.451&     15.385&     14.832&     14.329&      0.022&      0.010&      0.010&      0.011&      0.045&      0.049&      0.054\\
V102&  312.70291&   44.34814&     10.878&     19.077&     17.253&     16.048&     14.809&      0.079&      0.095&      0.076&      0.063&      0.405&      0.254&      0.245\\
V103&  312.73102&   44.29611&      9.383&     ---&     	18.679&     17.374&     15.866&      ---&      0.132&      0.112&      0.071&      0.509&      0.424&      0.242\\
V104&  312.71906&   44.27891&      8.446&     17.080&     15.437&     14.479&     13.550&      0.072&      0.043&      0.044&      0.036&      0.182&      0.161&      0.189\\
V105&  312.80222&   44.30257&      7.954&     20.071&     18.162&     16.995&     15.829&      0.078&      0.110&      0.104&      0.074&      0.523&      0.412&      0.288\\
V106&  312.75653&   44.26168&      7.359&     17.514&     15.951&     15.046&     13.905&      0.029&      0.187&      0.185&      0.150&      0.660&      0.631&      0.526\\
V107&  312.74460&   44.29190&      7.223&     17.905&     16.480&     15.422&     14.315&      0.175&      0.164&      0.150&      0.127&      0.603&      0.522&      0.435\\
V108&  312.71527&   44.37217&      7.176&     19.064&     17.630&     16.695&     15.951&      0.056&      0.051&      0.044&      0.042&      0.229&      0.189&      0.174\\
V109&  312.74783&   44.32752&      6.849&     20.560&     18.585&     17.256&     15.825&      0.408&      0.148&      0.117&      0.135&      0.609&      0.463&      0.515\\
V110&  312.70245&   44.35127&      6.217&     ---&     19.716&     18.147&     16.507&      ---&      0.333&      0.336&      0.393&      1.200&      1.417&      1.483\\
\hline
\end{tabular}}
\end{center}
{\footnotesize \textbf{Notes:} This table is available entirely in a machine-readable form in the online journal as supporting information. Only first-ten
lines are shown here for guidance regarding its form and content.}
\label{tab:per}
\end{table*}

The Lomb-Scargle periodogram \citep{lomb1976, scargle1982}, phase-dispersion minimization \citep{stellingwerf1978}, and analysis of variance \citep{schwarz1989} 
methods are used to find the periods for candidate variables. These different period determination algorithms allow us to ascertain the consistency of estimated periods. 
The period search was carried out in $10^{6}$ steps between 0.1 to 100 days. 
A typical uncertainty in the measured periods is found to be smaller than one tenth of a day.
A star is identified as periodic variable if the difference between periods in any 
two methods is smaller than $10^{-3}$ day. The remaining light curves are further visually inspected to select
variable candidates displaying periodic and/or non-periodic variations. In cases where the variability is not observed in all three filters, we restrict the sample to stars that display
at least $10\%$ of magnitude fluctuations and the amplitude is significantly larger than the dispersion in the light curve. 
The final sample of variables consists of 95 stars (see Fig.~\ref{fig:inst_sig}), out of which 56 show periodic variability in
at least one of the $VRI$ filters. Only 5 stars have amplitudes smaller than 0.1~mag in $I$-band. 
Multi-epoch photometric data of variable sources are tabulated in Table~\ref{tab:phot}. From {\it Gaia} DR2, 91 (out of 95) variables have parallaxes and proper motions and 
2 of those are fainter than 20~magnitude in $G$-band. 

The variable YSO sample with long-term time-series photometry in IC5070 is very limited \citep[][as discussed in Section 1]{findeisen2013, poljan2014, Ibryamov2018, froebrich2018}.
Therefore, a number of our targets are new variable sources identified in the IC 5070. Out of 77 stars common with \citet{rebull2011} in our target region, only 42 exhibit 
variability in the final sample. The photometric properties of periodic and non-periodic variables 
are listed in Table~\ref{tab:per}. The rms scatter for each variable source is added in quadrature to the photometric uncertainties listed in Table ~\ref{tab:per}.
Fig.~\ref{fig:lcs} shows the representative light curves of periodic and non-periodic variables. 

\section{Classification and evolutionary stages of the variable stars}
\label{sec:class}

In order to classify the variable sources, it is essential to find out their association with the Pelican Nebula region.
The color-color, color-magnitude diagrams and the kinematics information are very useful in identifying stars associated with a cluster \citep{panwar2014, lata2016}.

\subsection{Membership}
\begin{figure}
\centering
\includegraphics[width=0.5\textwidth]{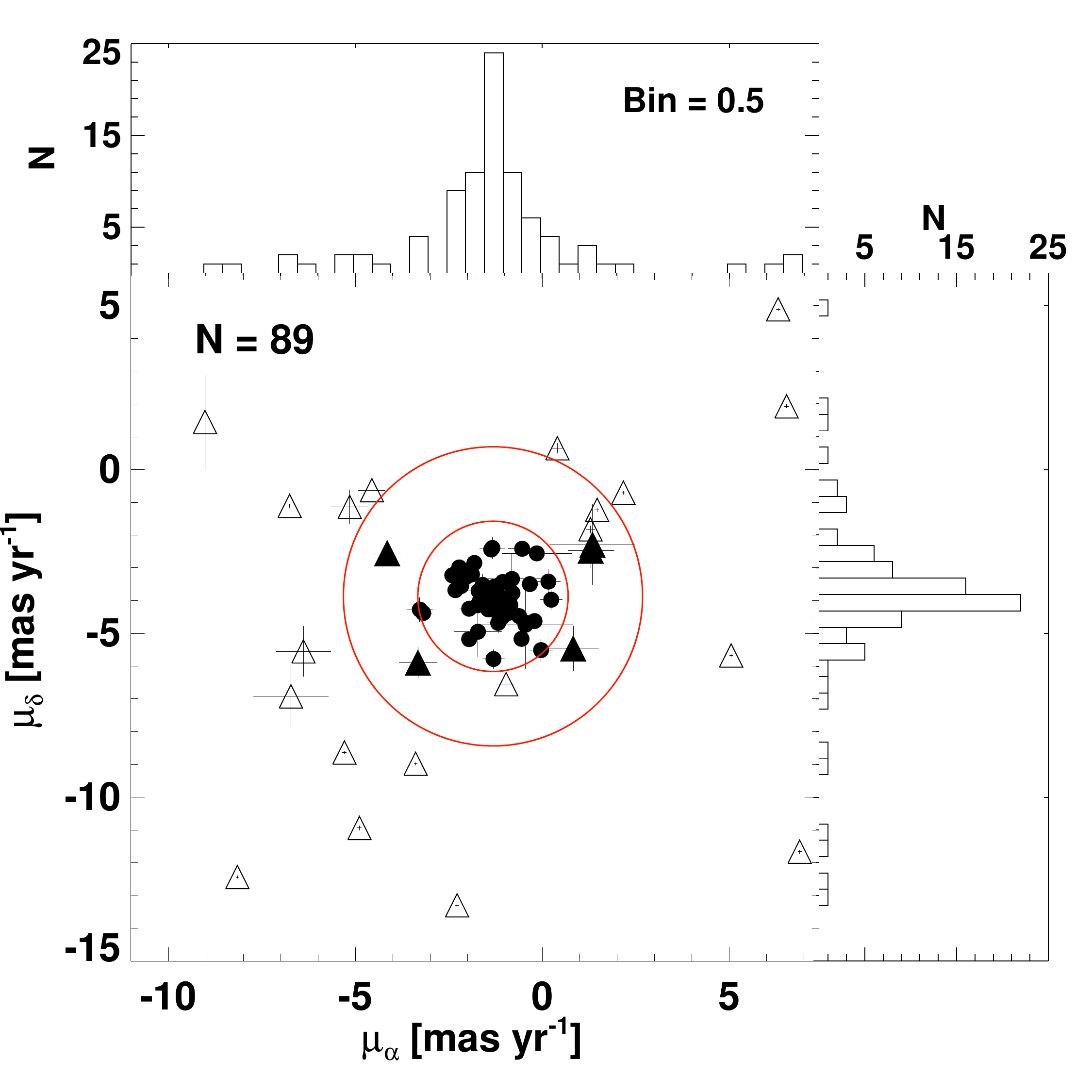}
\caption{The scatterplot of variable sources in the proper motion plane. Histograms for proper motions along right ascension and declination are also shown. 
The center of the circles is at the peak values of proper motions from the Gaussian distribution fits to the histograms. Inner/outer radius is of 2/4 mas per year. Filled circles
denote kinematically selected members. Open triangles represent $3\sigma$ ($\sim 2$ mas per year) outliers from the mean of the Gaussian distribution fits to the histograms of the 
proper motions. Filled triangles display variables for which the proper motions are consistent with mean value given their $3\sigma$ uncertainties. The bin-size used in the histograms and 
the number of stars shown are indicated on the top.}
\label{fig:pms}
\end{figure}

The parallaxes and proper motions from Gaia DR2 are used for variable sources to identify possible members of the Pelican Nebula region. 
Fig.~\ref{fig:pms} shows the scatterplot and histograms of proper motions for variable candidates. The Gaussian distribution fits to the proper motions 
along right ascension ($\mu_\alpha$) and declination ($\mu_\delta$) provide a mean value of -1.32 and -3.87 mas per year with a half width at half maximum 
of 0.65 and 0.67 mas per year, respectively. A distinct clustering in the scatterplot is visible around $\mu_\alpha\sim-1$~mas per year and $\mu_\delta\sim-4$~mas per year.
The inner circle is centered at the peak values from the Gaussian distribution with a radius of 2 mas per year equivalent to three times their half width at the half maximum.
Assuming the center of the radius as the proper motion of the cluster, 27 variables are found with proper motions beyond 3$\sigma$ of the mean values. Out of those, 5 stars have
large uncertainties in their proper motions which are sufficient to bring those within the inner radius. Therefore, 22 (out of 95) variables are likely MS or field stars 
and do not belong to the Pelican Nebula region. In addition, 4 stars do not have kinematic information available from {\it Gaia} data.

\begin{figure}
\centering
\includegraphics[width=0.5\textwidth]{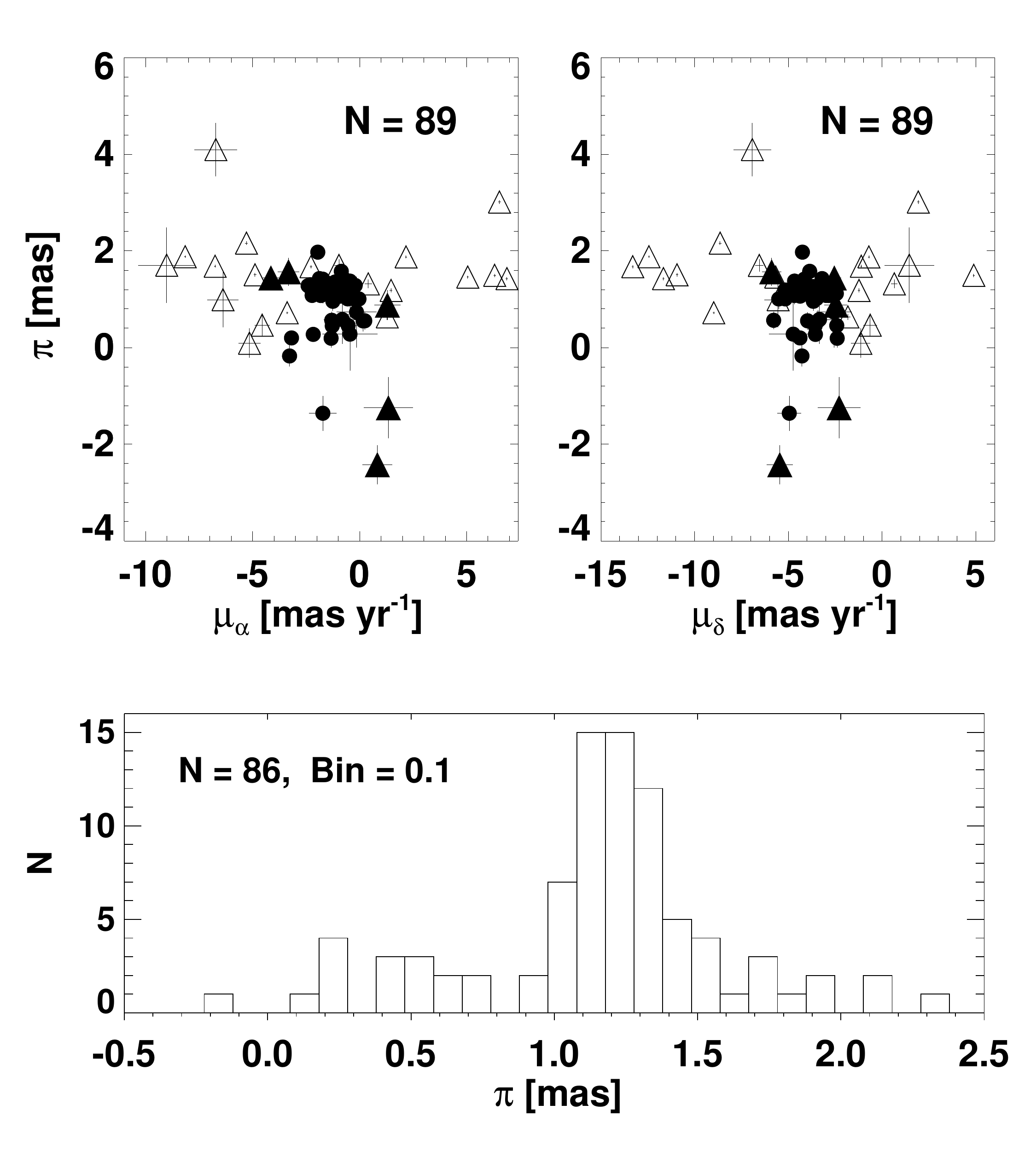}
\caption{{\it Top panel:} The scatterplot of parallaxes against the proper motions along right ascension and declination. Symbols are same as in Fig.~\ref{fig:pms}. 
{\it Bottom panel:} Histogram of parallaxes for variable sources from the Gaia DR2. The bin-size used in the histogram and the number of stars 
shown in the plots are indicated on the top.}
\label{fig:pms_radec}
\end{figure}

Fig.~\ref{fig:pms_radec} displays the scatterplot of parallaxes against the proper motions for variable candidates. In the parallax histogram, Gaussian distribution peaks 
at 1.21 mas with a half width at half maximum of 0.15 mas, corresponding to a median distance of $826.5$ pc with $1\sigma$ standard deviation of 101.6 pc, for all the 
target variables in our analysis. Four stars have negative parallaxes and the distances for 82 (out of 91) variables are consistent with the mean value given their uncertainties.
To estimate a robust distance to the IC 5070, we iteratively exclude stars that have kinematics and distances beyond $3\sigma$ of their mean values. The stars with excess astrometric noise 
more than 2 mas are also excluded from this analysis. For the remaining sample of 59 stars, individual distances and their associated low/high uncertainties are used from 
the catalog of \citet{coryn2018} to perform bootstrapping. We create 10$^4$ random realizations by perturbing the uncertainties in each iteration and finally fit a Gaussian 
distribution to estimate a distance of $857.5\pm 55.8$~pc to the IC 5070. The distance estimates to the Pelican Nebula region vary from 500 pc to over 1 kpc 
but a distance closer to 600 pc is preferred with a typical uncertainty of $10\%$ \citep{laugalys2007, reipurth2008, gui2009}. This commonly adopted distance 
is based on the extensive multi-color photometry that is used to determine color excesses, extinction and distances for hundreds of stars towards the Pelican 
Nebula region \citep[][and references within]{laugalys2007}.

\subsection{Color-magnitude and color-color diagrams}

\begin{figure}
\centering
\includegraphics[width=0.5\textwidth]{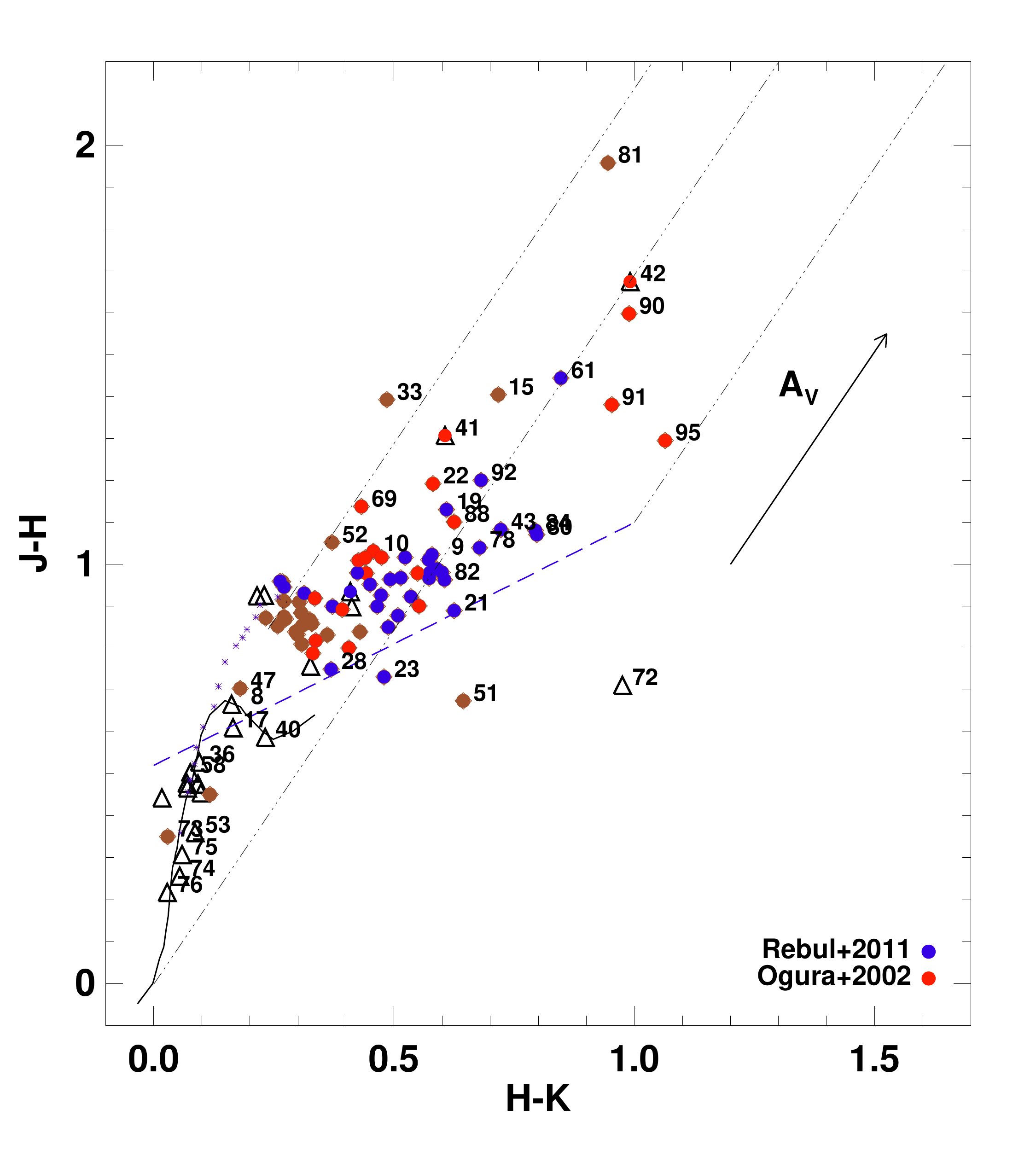}
\caption{Near-infrared color-color diagram for all variables. The solid curve and dotted curve represent the sequence of
dwarf and giants from \citet{bessell1988}. The locus of CTT stars is shown as long dashed lines \citep{meyer1997} while the dotted-dashed lines represent the reddening
vectors \citep{cohen1981}. Diamonds and open triangles represent the kinematic members and outliers, respectively. The solid arrow indicates reddening vector corresponding to $A_V=5$~mag.
The stars are numbered with the last two digits of their Star ID from Table~\ref{tab:per}.} 
\label{fig:tcds}
\end{figure}

Optical color-magnitude and near-infrared color-color diagrams are used to classify our variable candidates. The 2MASS $JHK_s$ data are available for 93 stars while for the remaining 
two stars, photometry are adopted from the UKIDSS Galactic Plane Survey \citep{lucas2008}. The 2MASS photometry is transformed to the California Institute of Technology (CIT) system 
using the relations provided on their website{\footnote{\tiny \url{http://www.astro.caltech.edu/~jmc/2mass/v3/transformations}}} to compare with the evolutionary models. Fig.~\ref{fig:tcds} 
represents the $J-H$/$H-K$ color-color diagram based on 2MASS data, typically used to classify the YSOs. The YSOs from \citet{rebull2011} and
\citet{ogura2002} are over-plotted in colored symbols. The sequence of dwarf and giants from \citet{bessell1988}, and the intrinsic locus of CTT stars \citep{meyer1997}
are also over-plotted. The three parallel lines are the reddening vectors drawn from the tip of the giant branch (left), from the base of the MS branch (middle) and from the tip of the intrinsic CTTSs
line (right). The extinction ratios to derive these reddening vectors are, $\frac{A_{J/H/K}}{A_V} = 0.265/0.155/0.090$, adopted from \citet{cohen1981}. 
In general, CTTSs with smaller near-infrared excess, WTTSs and field stars (MS/giants) occupy the region between the left and middle reddening vectors. Fig.~\ref{fig:tcds} 
shows that most of the variables that are outliers in the proper motions and lie below the intrinsic CTTSs locus are the MS stars. Two variables (V173 and V177) are members based on their 
proper motions but fall below the giant sequence. One of those, V173, does not have kinematic information. 
The CTTSs with large infrared excess are located in the region between the middle and right reddening vectors, while more moderate CTTSs with smaller infrared excess can also populate 
the region between left and middle reddening vectors, mixed with reddened WTTSs just above the CTT locus.
Some contamination is expected depending on the reddening, IR excess and also due to variability in single-epoch measurements.

\begin{figure}
\centering
    \includegraphics[width=.5\textwidth]{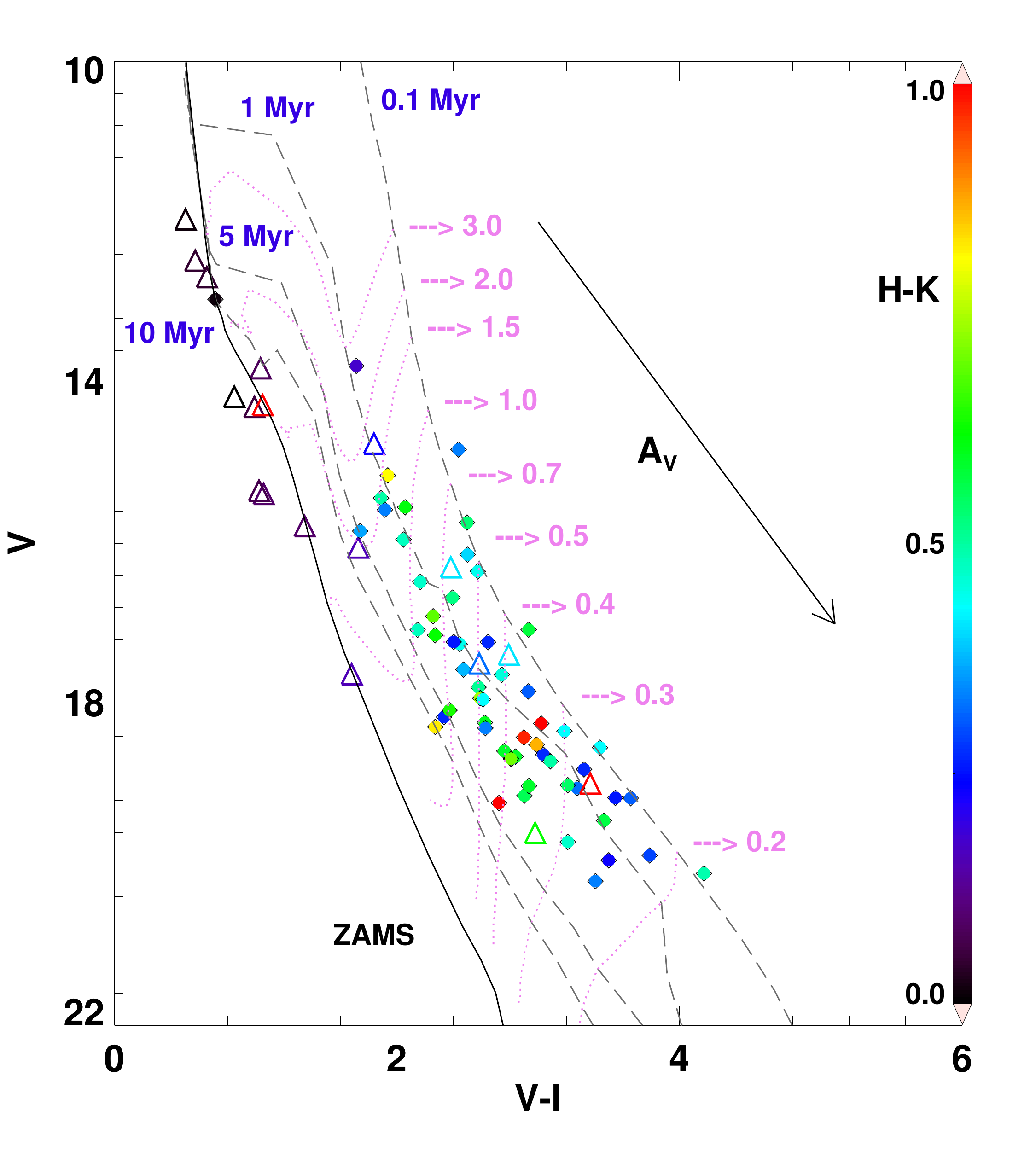}
 \caption{The apparent optical $V/V-I$ color-magnitude diagram for the variable candidates. The dashed curve shows the isochrones for PMS stars with different ages from \citet{siess2000}
while the solid curve is the ZAMS by \citet{girardi2002}. The dotted curves display the evolutionary tracks of PMS stars with different stellar masses. The isochrones and the evolutionary 
tracks are offset with respect to the average distance (857.5~pc) and extinction ($A_V=1.9$~mag). The color bar represents the 2MASS ($H-K$) color. 
Open triangles represent the kinematic outliers. The arrow indicates the direction of the reddening vector corresponding to $A_V=5$~mag.} 
 \label{fig:opt_cmd}
\end{figure}

\begin{figure}
\centering
    \includegraphics[width=.5\textwidth]{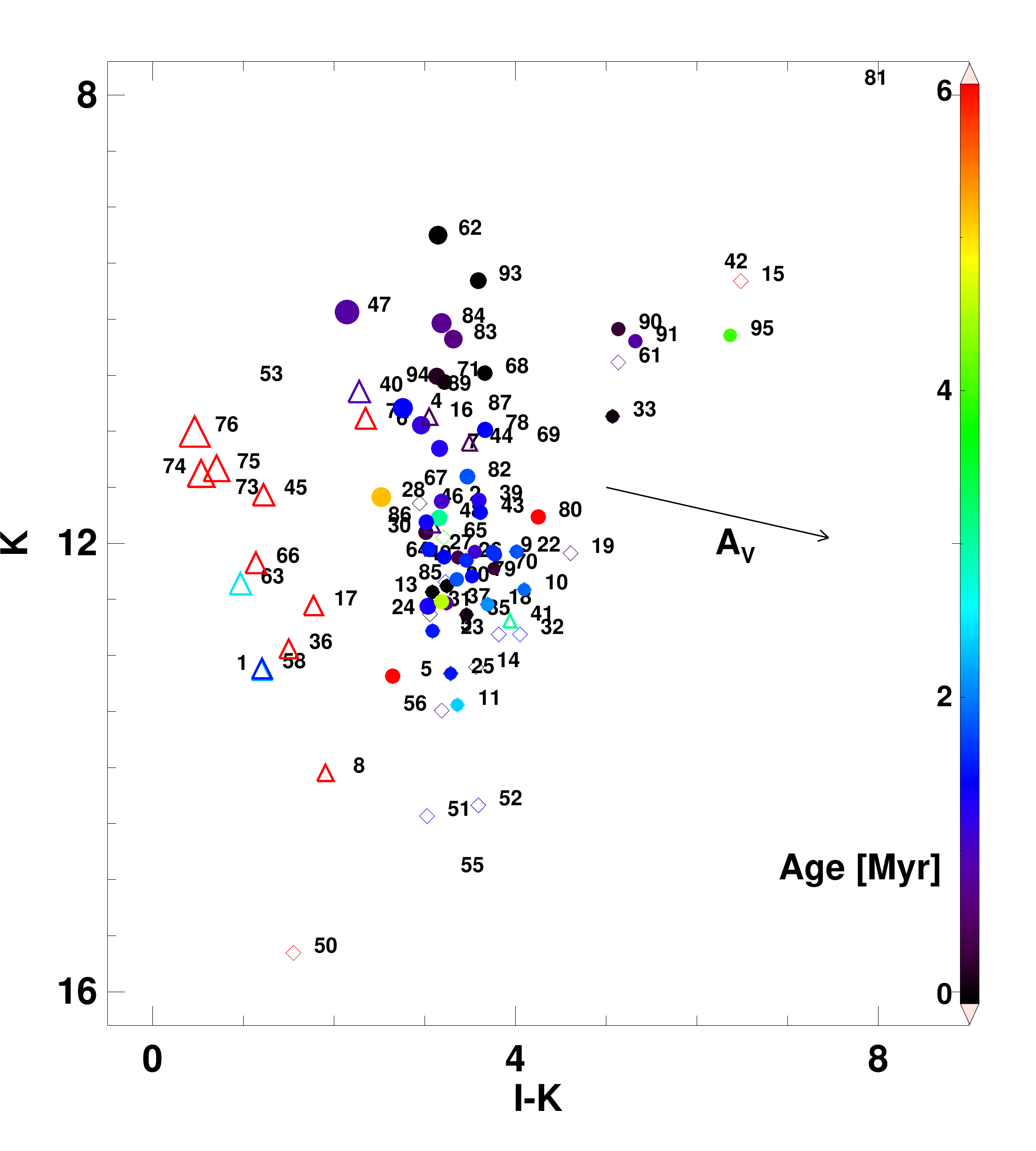}
 \caption{ $K/I-K$ color-magnitude diagram. The stars are numbered with the last two digits of their Star ID from Tables~\ref{tab:per}.  
The color bar represents the age determined from the isochrone-fitting. Open triangles represent the kinematic outliers.
The arrow indicates reddening vector corresponding to $A_V=5$~mag. The increase in symbol size represents the increase in mass obtained from
the evolutionary tracks.}
 \label{fig:ir_ccd}
\end{figure}

The optical $V/V-I$ color-magnitude diagram is a useful tool to ascertain the evolutionary status of variables and their probable cluster membership. Fig.~\ref{fig:opt_cmd} shows the
color-magnitude diagram for the variable candidates. The theoretical isochrones of 0.1, 1, 5 and 10 Myr age and evolutionary tracks of various masses for PMS stars
are plotted from \citet{siess2000}. We have also plotted the ZAMS with the solar metal-abundance (Z=0.02) from \citet{girardi2002}. The isochrones and evolutionary tracks are corrected 
for a distance of 857.5~pc to the cluster and for the average extinction of $A_V=1.9$~mag. The average extinction is estimated by tracing back the location of all variables
on the near-infrared color-color diagram to the intrinsic locus along the reddening vector. $A_V$ varies between zero to ten mag for these variables. 
The reddening towards the young stars is often differential across the star-forming region and a larger spread in reddening is expected throughout the cluster 
as shown in the extinction-maps by \citet{cambresy2002} based on 2MASS colors. 
The individual $A_V$ measurements may suffer from the possible degeneracies when the spectral information is not available, and in some cases the adopted average 
extinction may also be offset with respect to optical colors. The extinction measurements may differ from NIR estimates even if the spectral type is known \citep{herczeg2014}.
However, the average extinction based on near-infrared colors displays a reasonable fit to the PMS population 
in the optical color-color diagrams in our sample. Again, most of the proper motion outliers appear along the ZAMS in 
Fig.~\ref{fig:opt_cmd} except 5 variables (V116, V117, V140-V142) that follow younger PMS population. Light curves in $V$-band for V141 and V142 are of poor quality and it is possible that
either the color or magnitude for all these objects are spurious, leading to offset towards younger population.

\begin{table}
\caption{Age and mass estimates of variable stars based on isochrone-fitting.}
\centering
\scalebox{0.88}{
\begin{tabular}{c r r r r}
\hline
ID &	Mass ($M_\odot$) & Age (Myr)  	& Mass ($M_\odot$)   & Age (Myr)  \\
\hline
   & \multicolumn{2}{l}{~~~~~with mean-distance} & \multicolumn{2}{c}{with parallax distance}\\
\hline
V101& 1.54$\pm$ 0.05& 2.25$\pm$ 0.52& 1.42$\pm$ 0.05& 0.65$\pm$ 0.15\\
V102& 0.82$\pm$ 0.05& 1.16$\pm$ 0.15& 0.82$\pm$ 0.05& 1.63$\pm$ 0.30\\
V103& 0.57$\pm$ 0.05& 1.66$\pm$ 0.38& 0.55$\pm$ 0.05& 1.96$\pm$ 0.45\\
V104& 1.61$\pm$ 0.07& 1.54$\pm$ 0.36& 1.55$\pm$ 0.06& 1.95$\pm$ 0.28\\
V105& 0.79$\pm$ 0.06& 6.19$\pm$ 1.71& 0.80$\pm$ 0.07& 4.84$\pm$ 1.14\\
V106& 1.32$\pm$ 0.17& 1.28$\pm$ 0.69& 1.26$\pm$ 0.14& 1.48$\pm$ 0.86\\
V107& 1.10$\pm$ 0.15& 1.38$\pm$ 0.57& 1.05$\pm$ 0.09& 1.75$\pm$ 0.77\\
V108& 0.99$\pm$ 0.05& 45.00$\pm$ 9.36& 0.96$\pm$ 0.05& 45.00$\pm$ 9.58\\
V109& 0.58$\pm$ 0.05& 1.77$\pm$ 0.38& 0.51$\pm$ 0.05& 3.53$\pm$ 1.05\\
V110& 0.40$\pm$ 0.06& 1.98$\pm$ 1.45& 0.41$\pm$ 0.08& 1.78$\pm$ 1.16\\
V111& 0.35$\pm$ 0.05& 2.44$\pm$ 0.61& 0.36$\pm$ 0.05& 1.99$\pm$ 0.35\\
V112& 0.48$\pm$ 0.06& 0.16$\pm$ 0.04& 0.47$\pm$ 0.07& 0.16$\pm$ 0.03\\
V113& 0.53$\pm$ 0.05& 0.11$\pm$ 0.03& 0.57$\pm$ 0.05& 0.10$\pm$ 0.02\\
V116& 0.98$\pm$ 0.14& 0.65$\pm$ 0.50& 0.67$\pm$ 0.09& 17.77$\pm$ 8.09\\
V117& 1.46$\pm$ 0.05& 7.13$\pm$ 1.64& 0.91$\pm$ 0.05& 45.00$\pm$ 9.86\\
V118& 0.53$\pm$ 0.05& 2.24$\pm$ 0.53& 0.57$\pm$ 0.05& 1.71$\pm$ 0.22\\
V122& 0.55$\pm$ 0.05& 1.96$\pm$ 0.87& 0.59$\pm$ 0.05& 1.42$\pm$ 0.48\\
V125& 0.38$\pm$ 0.05& 1.61$\pm$ 0.67& 0.35$\pm$ 0.05& 1.95$\pm$ 0.43\\
V126& 0.54$\pm$ 0.07& 0.53$\pm$ 0.14& 0.56$\pm$ 0.07& 0.59$\pm$ 0.13\\
V127& 0.79$\pm$ 0.05& 1.62$\pm$ 0.21& 0.76$\pm$ 0.05& 1.76$\pm$ 0.25\\
V128& 1.56$\pm$ 0.05& 5.16$\pm$ 1.19& 1.54$\pm$ 0.05& 4.63$\pm$ 1.07\\
V130& 0.78$\pm$ 0.05& 0.45$\pm$ 0.07& 0.73$\pm$ 0.05& 0.49$\pm$ 0.06\\
V131& 1.06$\pm$ 0.74& 1.47$\pm$ 0.55& 1.05$\pm$ 0.74& 1.65$\pm$ 0.66\\
V133& 0.43$\pm$ 0.07& 0.25$\pm$ 0.05& 0.44$\pm$ 0.05& 0.10$\pm$ 0.02\\
V135& 0.42$\pm$ 0.05& 0.27$\pm$ 0.06& 0.36$\pm$ 0.06& 0.57$\pm$ 0.10\\
V136& 1.39$\pm$ 0.05& 33.39$\pm$ 7.69& 1.37$\pm$ 0.05& 31.49$\pm$ 7.25\\
V137& 0.58$\pm$ 0.05& 0.90$\pm$ 0.42& 0.57$\pm$ 0.05& 0.92$\pm$ 0.40\\
V138& 0.82$\pm$ 0.05& 1.01$\pm$ 0.23& 1.02$\pm$ 0.19& 0.63$\pm$ 0.05\\
V139& 0.77$\pm$ 0.13& 1.33$\pm$ 1.63& 1.04$\pm$ 0.15& 0.16$\pm$ 0.14\\
V140& 1.86$\pm$ 0.08& 0.98$\pm$ 0.22& 2.27$\pm$ 0.13& 0.44$\pm$ 0.10\\
V141& 0.47$\pm$ 0.06& 3.13$\pm$ 1.60& 0.37$\pm$ 0.09& 7.21$\pm$ 1.66\\
V143& 0.72$\pm$ 0.08& 1.56$\pm$ 1.24& 0.71$\pm$ 0.07& 1.59$\pm$ 1.25\\
V144& 0.77$\pm$ 0.06& 0.57$\pm$ 0.07& 0.69$\pm$ 0.05& 1.19$\pm$ 0.27\\
V145& 1.85$\pm$ 0.05& 12.51$\pm$ 2.88& 1.60$\pm$ 0.05& 3.41$\pm$ 0.69\\
V147& 2.38$\pm$ 0.12& 0.94$\pm$ 0.31& 2.39$\pm$ 0.10& 1.04$\pm$ 0.38\\
V148& 1.03$\pm$ 0.15& 3.15$\pm$ 2.98& 1.03$\pm$ 0.15& 3.20$\pm$ 3.05\\
V149& 0.65$\pm$ 0.05& 1.76$\pm$ 0.22& 0.66$\pm$ 0.05& 1.67$\pm$ 0.21\\
V158& 1.56$\pm$ 0.05& 1.63$\pm$ 0.38& 1.46$\pm$ 0.05& 0.61$\pm$ 0.07\\
V159& 0.57$\pm$ 0.05& 1.16$\pm$ 0.16& 0.59$\pm$ 0.05& 1.48$\pm$ 0.34\\
V162& 1.39$\pm$ 0.05& 0.10$\pm$ 0.02& 3.07$\pm$ 0.14& 0.10$\pm$ 0.02\\
V163& 1.85$\pm$ 0.05& 2.54$\pm$ 0.59& 1.73$\pm$ 0.05& 0.85$\pm$ 0.11\\
V164& 0.72$\pm$ 0.05& 1.50$\pm$ 0.22& 0.60$\pm$ 0.05& 2.26$\pm$ 0.52\\
V166& 1.74$\pm$ 0.05& 9.38$\pm$ 2.62& 1.46$\pm$ 0.05& 0.41$\pm$ 0.06\\
V168& 0.83$\pm$ 0.05& 0.10$\pm$ 0.02& 0.82$\pm$ 0.05& 0.28$\pm$ 0.10\\
V170& 0.41$\pm$ 0.07& 0.40$\pm$ 0.22& 0.45$\pm$ 0.08& 0.34$\pm$ 0.19\\
V171& 1.10$\pm$ 0.08& 0.34$\pm$ 0.09& 0.97$\pm$ 0.05& 5.50$\pm$ 0.11\\
V172& 1.71$\pm$ 0.05& 22.71$\pm$ 5.23& 1.56$\pm$ 0.05& 2.41$\pm$ 0.56\\
V174& 2.75$\pm$ 0.08& 46.59$\pm$ 7.07& 2.39$\pm$ 0.05& 18.97$\pm$ 8.38\\
V175& 2.56$\pm$ 0.06& 26.70$\pm$ 4.15& 2.13$\pm$ 0.05& 3.09$\pm$ 0.27\\
V176& 3.33$\pm$ 0.11& 35.08$\pm$ 9.10& 2.26$\pm$ 0.05& 5.24$\pm$ 0.10\\
V178& 0.96$\pm$ 0.10& 1.55$\pm$ 0.98& 0.96$\pm$ 0.10& 1.54$\pm$ 1.00\\
V179& 0.59$\pm$ 0.05& 1.54$\pm$ 0.61& 0.61$\pm$ 0.05& 1.25$\pm$ 0.41\\
V180& 0.80$\pm$ 0.16& 9.82$\pm$ 6.42& 0.83$\pm$ 0.16& 4.93$\pm$ 3.67\\
V182& 0.89$\pm$ 0.08& 1.95$\pm$ 0.69& 0.87$\pm$ 0.10& 2.17$\pm$ 0.85\\
V183& 1.36$\pm$ 0.27& 0.80$\pm$ 0.74& 1.34$\pm$ 0.20& 0.97$\pm$ 0.90\\
V184& 1.61$\pm$ 0.22& 0.91$\pm$ 0.56& 1.53$\pm$ 0.26& 0.99$\pm$ 0.63\\
V185& 0.65$\pm$ 0.13& 1.92$\pm$ 1.80& 0.60$\pm$ 0.11& 2.34$\pm$ 2.25\\
V186& 0.83$\pm$ 0.05& 1.42$\pm$ 0.31& 0.85$\pm$ 0.05& 1.52$\pm$ 0.37\\
V188& 0.77$\pm$ 0.12& 4.62$\pm$ 6.49& 0.77$\pm$ 0.14& 5.78$\pm$ 7.52\\
V190& 0.62$\pm$ 0.13& 0.44$\pm$ 0.31& 0.66$\pm$ 0.11& 0.39$\pm$ 0.27\\
V191& 0.64$\pm$ 0.11& 0.94$\pm$ 0.94& 0.63$\pm$ 0.10& 1.03$\pm$ 0.99\\
V192& 0.59$\pm$ 0.05& 1.72$\pm$ 1.18& 0.65$\pm$ 0.06& 1.30$\pm$ 0.63\\
V193& 1.05$\pm$ 0.13& 0.10$\pm$ 0.06& 1.11$\pm$ 0.06& 0.30$\pm$ 0.19\\
V194& 0.87$\pm$ 0.17& 0.25$\pm$ 0.17& 1.49$\pm$ 0.10& 0.10$\pm$ 0.02\\
V195& 0.52$\pm$ 0.08& 4.08$\pm$ 2.18& 0.43$\pm$ 0.08& 16.85$\pm$ 7.40\\
\hline
\end{tabular}}
\label{tab:age_mass}
\end{table}

The age and mass of variable candidates are estimated by comparing their locations in the observed color-magnitude 
diagram with the theoretical PMS isochrones of \citet{siess2000} and \citet{bressan2012}. These estimates are likely to be uncertain due to several obvious 
reasons, for example, isochrones uncertainties, lack of precise reddening corrections, binarity and variability effects on colors.
In order to provide an accurate range of age and mass estimates, a single distance of $857.5\pm 55.8$ pc and an extinction of $A_V=1.9$~mag is adopted 
to offset absolute magnitudes. The isochrones and evolutionary tracks on the color-magnitude diagram are interpolated using a 2-dimensional grid. 
Firstly, the unevenly-spaced isochrones and evolutionary tracks are trianglulated to form a regular-grid. Then the age and mass estimates are
obtained for each star on the observed color-magnitude diagram by interpolating the 2-dimensional regular grid using inverse distance weighted interpolation i.e. 
the nearest age and mass grid to the data points in the color-magnitude diagram are given higher weights.
In order to be more robust, the errors in photometry, distances and reddening  are added in quadrature to perform a bootstrapping by allowing the magnitude and colors to change 
within $1\sigma$. Several ($10^4$) random realizations are created and the median values of the age and mass are adopted as final estimates while the standard deviation as the 
error on the adopted values. Table~\ref{tab:age_mass} lists the age and mass estimates from the isochrone-fitting to the color-magnitude diagram for those variables 
that have both $V$ \& $I$ measurements and accurate parallaxes. Most stars have masses $\le 2 M_\odot$ and ages less than 10 Myr 
with a median uncertainty of $9\%$ and $24\%$ in mass and age estimates, respectively.
Additionally, individual distances from the catalog of \citet{coryn2018} are used to estimate the absolute $V$-band magnitude for our variables and perform the same analysis. 
From Table~\ref{tab:age_mass}, the difference in the masses and ages between these two approaches are typically smaller than their quoted 
uncertainties. However, masses for 7 stars and the ages for 10 stars in our sample differ by more than $3\sigma$ of their uncertainties in the single cluster distance approach.

Theoretical isochrones from \citet{bressan2012} are also used for a systematic comparison and a median difference of $\sim 16\%$ is noted for masses greater 
than $0.6 M_\odot$. However, low-mass ($\le 0.6 M_\odot$) evolutionary tracks from \citet{bressan2012} display a systematic offset with respect to \citet{siess2000} and the median
difference in the estimated masses is around $29\%$. Similar differences are also found in age estimates for younger than 1 Myr population but the two set of age estimates from different
theoretical isochrones are consistent given their large uncertainties. Typically the masses inferred for individual stars from theoretical models are accurate to better than $10\%$
for masses $> 1 M_\odot$ but highly discrepant for sub-solar masses \citep{stassun2014, david2019}. Similarly, the systematic differences between ages predicted with different 
theoretical isochrones increase towards younger ages and for lower masses \citep{hillenbrand2008, soderblom2014}. With adopted distance and extinction to the IC 5070 cluster, 
a median mass and age of $\sim 0.82M_\odot$ and $1.55$ Myr is found based on the isochrone fitting to the observed color-magnitude diagram.
Fig.~\ref{fig:ir_ccd} displays $K/I-K$ color-magnitude diagram for variable sources with different stellar masses and ages where the MS or field variables are distinctly separated 
from younger PMS stars.

\begin{table*}
\caption{Comments on the classification and variability of the light curves.}
\scalebox{0.87}{
\begin{tabular}{l l p{7cm} p{9.5cm}}
\hline
ID & Type & Comments on classification & Comments on variability  \\
\hline
V101 &     MS/Field		& Based on kinematics, CMD and CCD		 & Low-amplitude $\sim 0.05$ mag in $VRI$				\\
V102 &  	WTT		& No NIR excess, PMS in CMD and CCD 	 	 & Strong periodicity perhaps due to spots, $\Delta I \sim 0.25$ mag		\\	
V103 &  	WTT		& No IR excess, PMS in CMD and CCD 		 & Periodic variability, $\Delta I \sim 0.25$ mag				\\	
V104 &  	CTT		& On CTT locus in NIR CMD, MIR excess		 & Small-amplitude ($< 0.2$~mag) brightening events and multiple extinction dips, periodic \\ 	
V105 & 		WTT		& No IR excess, PMS in CMD, near tip of the giant branch	  & Strong periodicity in multiple cycles, variability due to spots\\
V106 & 		CTT		& Above CTT locus in CMD, MIR excess		 & Periodic with a brighter secondary minima, possible occultations and sharp drop of 0.5 mag in $I$\\ 
V107 & 		WTT		& PMS in CMD, No NIR excess		   	 & Periodic variability with large scatter in light curve, two sequences around minima \\	
V108 & 		MS/field	& Kinematic outlier, In CMD and CCD		 & Small-amplitude variable, $\Delta V \sim 0.2$ mag				\\	
V109 & 		CTT		& In CTT region in NIR CCD, MIR excess	  	 & Large scatter in light curve with a periodic signal \\
V110 & 		CTT		& MIR excess, PMS in CMD			 & Large amplitude $\sim 1.5$ mag in $I$, quasi-periodic amplitude variations	\\
V111 & 		CTT		& Close to CTT locus in CCD, MIR excess		 & Periodic variability, $\Delta I \sim 0.35$ mag	\\
V112 & 		WTT		& No IR excess, PMS in CCD and CMD	  	 & Small-amplitude periodic variable, $\Delta I \sim 0.2$ mag	\\
V113 & 		WTT		& PMS in CCD and CMD, small MIR excess	 	 & Periodic variability with several random epochs of fainter than median mag \\		
V114 & 		WTT		& PMS in CCD and CMD, close to the locus of CTT	 & Strong periodicity with $\Delta I \sim 0.5$ mag	\\
V115 & 		Field/MS	& Distance outlier, PMS with NIR excess but no MIR excess		 & Smaller number of epochs, observed only in $RI$ \\	
V116 & 		MS/field	& Kinematic and distance outlier, PMS in CMD and CCD 	 & Eclipsing feature in the phased light curve \\
V117 & 		MS/field	& Kinematic and distance outlier and in CCD, PMS in CMD	 & Low-amplitude $\sim 0.05$ mag in $I$		\\	
V118 & 		CTT		& Near the locus of CTT in CCD, MIR excess	 & Periodic variation with scatter in the light curve \\
V119 & 		CTT		& PMS in $RI$ CMD and in NIR CCD, IR excess	 & Periodicity with $\Delta I \sim 0.3$~mag \\						
V120 & 		WTT		& No IR excess, near the tip of giant branch 	 & Scatter in the periodic light curve \\	
V121 & 		CTT		& On CTT locus, large MIR excess		 & Possible accretion burst, Periodic signals with 0.5 mag in $RI$ \\  
V122 & 		CTT		& IR excess, PMS in CMD and CCD			 & Strong periodicity with $\Delta I \sim 1$ mag and evidence of extinction events \\
V123 & 		CTT		& Just below CTT locus, MIR excess		 & Periodic variation with scatter in the light curve \\						
V124 & 		WTT		& No IR excess, PMS in CMD	  		 & Periodicity with mall-amplitude $\sim 0.1$ mag in $I$ \\				
V125-V126 &	WTT		& In CMD, near tip of giant branch in CCD	 & Scatter in the phased light curves \\
V127 & 		CTT		& PMS in CMD and CCD, MIR excess		 & Periodic variation with scatter in the light curve \\	
V128 & 		WTT		& Near CTT locus in CCD, No IR excess		 & Possible extinction events, small-amplitude $\sim 0.12$ mag in $I$	\\
V129 & 	        WTT		& Close to the tip of dwarf branch, No IR excess & Weak periodicity with smaller amplitude, ($\Delta R \sim 0.3$ mag\\
V130 &  	WTT		& No IR excess, PMS in CMD and CCD		 & Periodic variations of a one tenth of the magnitude \\ 
V131 & 		WTT		& No IR excess, PMS in CMD, nead tip of the giant branch	& Strong periodicity \\		
V132 & 		CTT		& Large MIR excess, just above CTT locus in CCD	 &   	\\					
V133 & 		Field 		& In CCD, PMS in CMD, Not a distance/kinematic outlier 	 & Periodic light curve	\\
V134 & 		WTT 		& PMS in CCD, No IR excess			 & Weak periodic variations	\\
V135 & 		WTT		& PMS in CMD and CCD, No IR excess		 & Periodic in $I$ with scatter in $VR$-bands \\	
V136 & 		MS		& In CMD and CCD				 &  \\
V137 & 		WTT		& In CMD and CCD, No IR excess			 & Possible short-time extinction events		 	\\
V138 & 		Field		& PMS in CMD, Just above CTT locus in CCD, No IR excess	 & A detached eclipsing binary system	\\
V139 & 		CTT		& On CTT locus in CCD, MIR excess, PMS in CMD    & Several extinction dips of $\sim 1$~mag or more\\
V140 & 		MS		& In CCD, PMS in CMD				 & Small amplitude of 0.1 mag in $I$	\\
V141 & 		MS/Field	& Kinematic outlier, PMS in CMD and CCD 	 & $>1$ mag monotonic dip in $I$ over 50 days, Possible long-period variable \\
V142 & 		MS/Field	& Kinematic outlier, PMS in CMD			 & Multiple smaller brightening events in 2012 and a fading event in 2013 \\
V143 & 		CTT		& IR excess, in CTT region in CMD		 & Several fading and extinction dips of amplitude $\sim 1$ mag in $VI$ \\
V144 & 		Field		& Kinematic outlier, PMS in CMD and CCD, MIR excess		 & Low-amplitude and periodic variation with large scatter \\
V145 & 		MS		& Kinematic and distance outlier, In CMD and CCD 		 & Low-amplitude variable \\
V146 & 		WTT		& No NIR excess, near tip of the Giant branch	 & Near-sinusoidal light curve, an extinction event lasts 20 days in Nov 2012 \\
V147 & 		MS		& No IR excess, PMS in CMD			 & (Semi-)irregular variability \\ 
V148 & 		CTT		& PMS in CMD and CCD, MIR excess		 & Small brightening events in 2012 followed by long-term fading event in 2013	 \\
V149 & 		CTT		& In CTT region in CMD, MIR excess	   	 & Several small extinction dips of up to 0.3 mag, a possible burst in May 2013 \\
V150 & 		WTT		& Close to CTT locus in CCD			 & Periodic, Observed in $I$ only with amplitude $> 1$~mag			\\
V151 & 		CTT		& On the CTT locus in CCD, PMS in $RI$ CMD	 & $\sim 1$~mag burst followed by dipping for 20 days in May 2013	 \\
V152 & 		WTT		& In CCD, PMS in $RI$ CMD			 & Observed in $RI$-only \\												
V153-V155 & 	MS/field	& Along giant/dwarf sequence in CCD, no IR excess & Periodicity with scatter in the light curve \\
\hline
\end{tabular}}
\label{tab:class}
\end{table*}

\begin{table*}
\scalebox{0.87}{
\begin{tabular}{l l p{7cm} p{9.5cm}}
\hline
V156	 & 	WTT		& In CCD, no IR excess				 & Periodic but observed in single filter \\
V157 & 		CTT		& MIR excess, Below CTT locus in CCD, PMS in CMD & Fading and extinction events of $\sim 1$ mag in Oct and Dec 2012\\
V158 & 		MS		& Kinematic outlier, In CMD and CCD			 & Small-amplitude variable			 	\\	
V159 & 		CTT		& In CMD and CCD, Large MIR excess			 & Small fading event followed by a bursting event within 10 days in Nov 2012		  \\
V160 & 		MS		& Kinematic outlier, In CMD and CCD			 & 		 	\\
V161 & 		CTT		& High IR excess, PMS in CMD and CCD		 & Brightening event starting Dec 2012 lasts 20 days  \\
V162 & 		Field/MS	& Distance outlier, In CMD and CCD, No IR excess  		 &  \\	
V163 & 		MS		& Kinematic outlier, In CMD and CCD			 & \\
V164 & 		CTT	  	& In CMD and CCD, small MIR excess		 & Multiple small brightening events of $\sim 0.4$~mag and a burst in mid May 2013  \\	
V165 & 		CTT	  	& In CMD and CCD, MIR excess			 & Fading in Sep 2012 for 20 days followed by rise in magnitude for a month \\
V166 & 		MS		& Kinematic and distance outlier, In CMD and CCD 			 &				 	\\
V167 & 		CTT		& In CMD and CCD, MIR excess			 & Multiple brightening events in 2012 and fading in 2013, Two bursts in Dec 2012 and May 2013 \\
V168 & 		CTT		& Close to CTT region in CMD, MIR excess	 & 20 day extinction event in May 2013, One bursting event in Mar 2013 	\\			
V169 & 		CTT	 	& No NIR excess, high MIR excess		 & Two brightening events in Nov and Dec 2012, Fading in Apr 2013	\\
V170 & 		CTT		& PMS in CMD and CCD, MIR excess	  	 & Multiple fading and brightening between Nov 2012 to Apr 2013, A small burst in Mar 2013\\
V171 & 		CTT		& In CMD and CCD, MIR excess			 & Possible burst in late May 2013\\
V172-V177 &  	MS/field	& In CMD and CCD				 &		  \\
V178 & 		CTT	  	& Inside CTT region in CCD, MIR excess		 & Multiple brightening and extinction events, Coming out of bursts in Oct 2012 and Apr 2013  \\
V179 & 		CTT	  	& In CTT region in CMD, high MIR excess		 & Extinction event of over $1$~mag in Oct 2012 lasting for a month, Two bursting events in Jan 2013 lasting a week\\
V180 & 		CTT	  	& Just above CTT locus in CCD, MIR excess	 & Several fading and extinction events with amplitudes up to $\Delta I \sim 2$ mag \\
V181 &	   	CTT  		& High NIR excess, no MIR excess		 & Possible long-period variable, Fading in 2012, rise in 2013	\\
V182 & 		CTT	  	& High MIR excess, Inside CTT region in CCD	 & Brightening in Nov 2012 and a 10 day fading in Jan 2013 \\
V183 & 		CTT	  	& In CCD, MIR excess				 & Several brightening events and extinction events with amplitudes up to $\Delta I \sim 1$ mag \\
V184 & 		CTT	  	& Inside CTT region in CCD, MIR excess		 & 10 day Fading in Oct 2012 and June 2013, $~50$ day brightening started in Nov 2012 \\
V185 &		CTT		& In CMD and CCD, MIR excess			 & Multiple brightening and fading with amplitude $\sim 1.5$ mag in $I$, A burst in Nov 2012\\	
V186 &		CTT  		& In CMD and CCD, near tip of the giant branch   & Fading from brightest to faintest magnitude over 60 days from Dec 2012 \\		
V187 &  	CTT	  	& In CMD and CCD, small MIR excess		 & Multiple brightness dipping events in Oct 2012, Jan \& May 2013. \\
V188 &  	CTT	  	& MIR excess, Close to CTT region in CCD	 & Several extinction events as large as $\Delta I > 2$ mag\\
V189 & 		CTT		& In CMD and CCD, MIR excess			 & Multiple fading events of up to $\Delta I > 1.2$ mag, faintest in Jun 2013 \\
V190 & 		CTT	  	& High IR excess, Inside CTT region in CCD	 & Significant extinction event and rise in magnitude in May 2013 \\ 		
V191 & 		CTT	  	& IR excess, inside CTT region in CCD 		 & A brightening event in Apr 2013 with total amplitude of $\sim 1.5$ mag in $I$ \\ 					
V192 & 		CTT	  	& PMS in CMD and CCD, IR excess			 & Multiple dipping of $\sim 1.5$ mag in $I$, A burst in Dec 2012\\ 					
V193 & 		CTT	  	& In CMD and CCD, MIR excess		         & Significant fading from median mag in Oct/Dec 2012 and May 2013 \\	
V194 & 		CTT	  	& In CMD and CCD, MIR excess			 & Brightening in April 2013 of $\sim 0.4$ mag, Possible burst in Oct 2012 \\	
V195 & 		CTT		& NIR excess, inside CTT region	in CCD		 & Significant distinct brightening event of $1$ mag between Nov 2012 to Jan 2013 \\
\hline
\end{tabular}}
\end{table*}

\subsection{Classification of variables}

In addition to the proper motions, distances, color-color and color-magnitude diagrams, we use the light curve structure to classify the variables in different subclasses.
All the kinematic outliers (22 stars) are classified as MS/field stars. A few additional variables (6 stars including 2 distance outliers) are also classified as MS stars as they 
fall below the giant sequence on the near-infrared color-color magnitude diagram or follow the ZAMS in the optical color-magnitude diagram. The rest of the PMS stars are classified 
as class II/III sources based on their large/small IR excess. These Class II/III sources are further identified as T Tauri candidates based on their variability signatures. Class III objects 
with small or no infrared excess displaying periodicity and smaller amplitude variations are classified as WTTSs. Variability signatures in WTT candidates is
dominated by the asymmetric distribution of spots on the stellar surface. Conversely, Class II PMS stars with moderate to large infrared excess show significantly large 
magnitude fluctuations. Their variability features include either single or multiple fading and brightening events. Class II objects displaying these extinction or bursting events 
are classified as CTTSs. WTT candidates show strong-periodicity with rotation periods typically smaller than 10 days in our sample. These lie close to the CTT locus in $J-H/H-K$ color-color 
diagram and exhibit amplitudes smaller than 0.8 mag in $V$-band. CTT candidates show evidence of single or multiple fading/brightening and extinction events that last different 
time-spans. Several CTTSs also display quasi-periodic variability with variable amplitudes. The adopted classifications and their observed variability characteristics are 
listed in Table~\ref{tab:class} for all variables. The magnitude variations larger than 1 mag in $I$-band are seen in 16 variables in our sample; CTT candidates 
like V180 can exhibit amplitudes as large as 2.5 mag in $I$. The average amplitude variation for all PMS stars is $\sim 0.5$ magnitude in $I$. These variations are at least twice as 
large for CTTSs than for the WTTSs. Out of 95 variables, 28 stars are classified as MS or field and 67 stars as PMS stars (45 CTTSs and 22 WTTSs). Out of all variable
candidates, $60\%$ display clear periodicity in the light curves. Based on the signature of the variability, nearly $70\%$ of all PMS stars display distinct CTT behaviour 
with large magnitude fluctuations while only $\sim 30\%$ show strong periodicity seen in WTTSs.

\begin{figure*}
\centering
\includegraphics[width=1.0\textwidth]{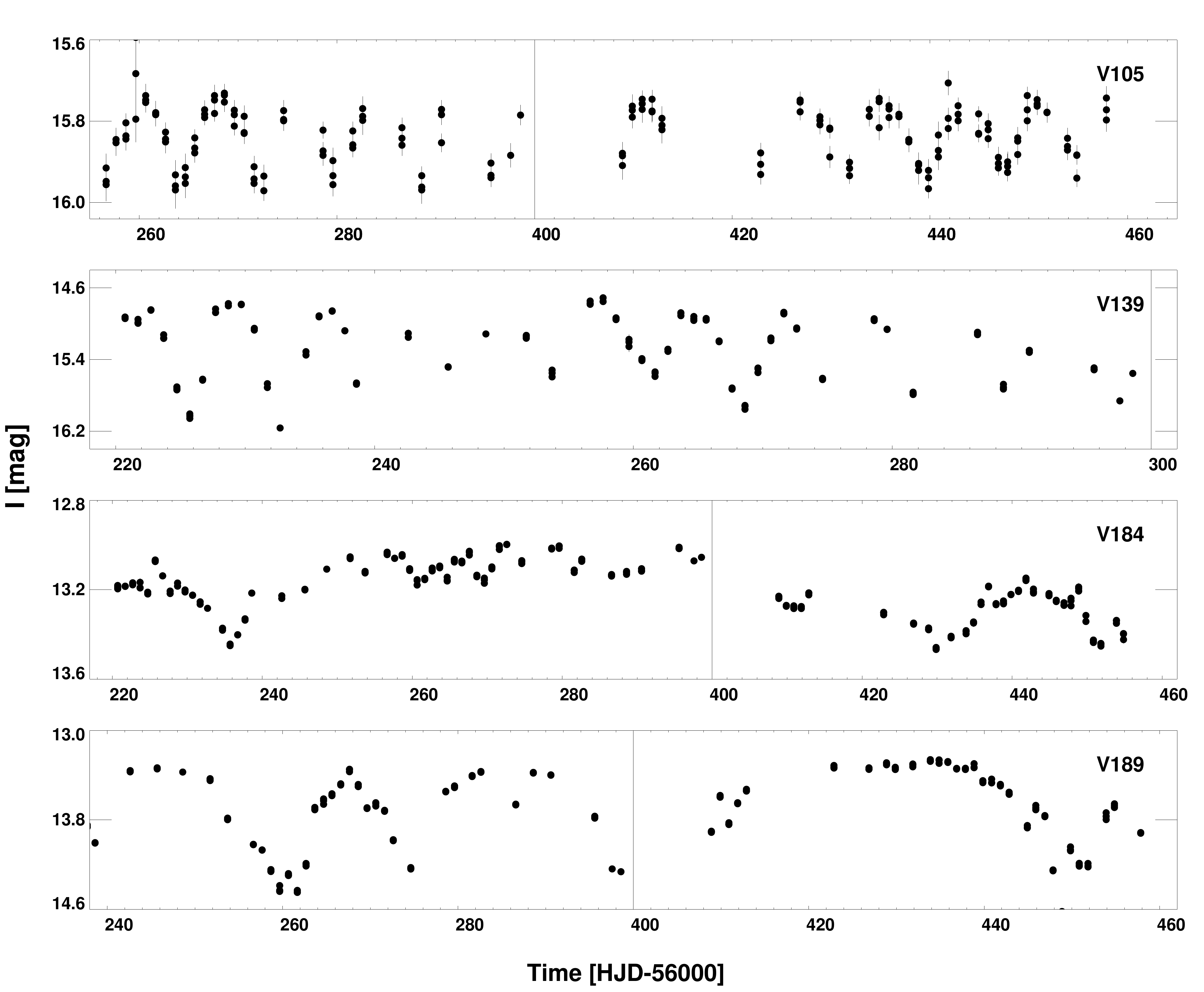}
\caption{The $I$-band light-curves of few TTSs in our sample. Top panel shows a WTT displaying periodic brightness variations with similar amplitudes. Second panel shows a T Tauri star with
a periodic variation together with multiple extinction events where magnitude fades significantly than its median value. Bottom two panels display CTTSs with different variability
signatures (see text for details).}
\label{fig:tt_lcs}
\end{figure*}

\begin{figure*}
\centering
\includegraphics[width=1.0\textwidth]{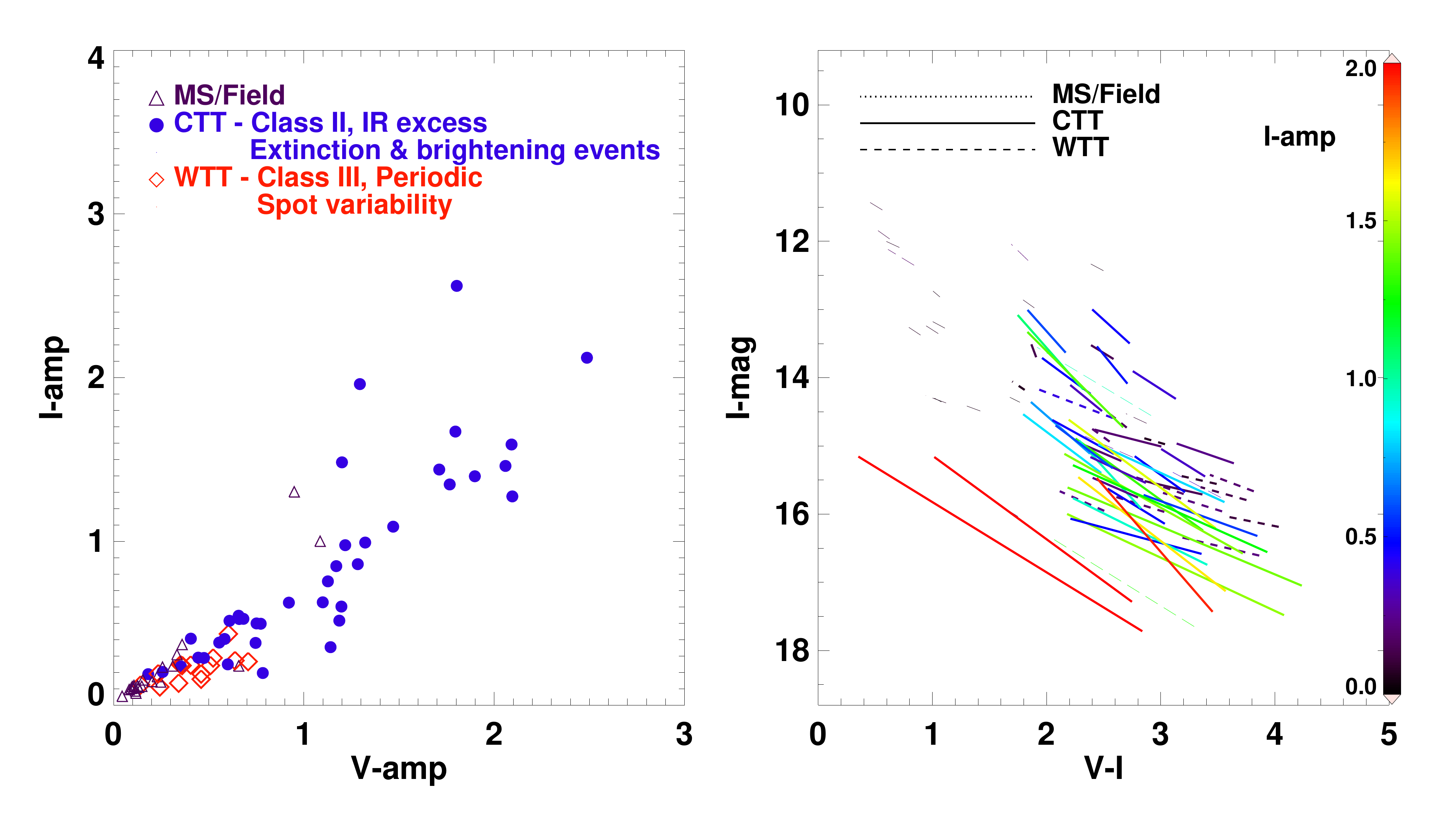}
\caption{{\it Left panel:} Variation of $V$ and $I$-band amplitudes for all variables. {\it Right panel:} Variability in the amplitude-color plane plotted in the color-magnitude
diagram. Variation along $I$-mag is the difference between maximum and minimum while ($V-I$) represents the range of the color curve. The color-bar denotes the range of $I$-band magnitude.}
\label{fig:amp_col}
\end{figure*}

\section{Pre-main sequence variables}
\label{sec:pms}

We discuss the physical and variable characteristics of the PMS stars in our sample in the following subsections.

\subsection{T Tauri variables}

Our sample of PMS variables includes 45 CTT and 22 WTT candidates. \citet{herbst1994} have shown that the 
typical variation in $V$-band amplitudes for WTTSs is up to three tenths of a magnitude with extreme values approaching 1 mag. These variations often occur 
within a period range of 0.5-18 days. WTT candidates in our sample have periods up to 10 days and a median
$V$-band amplitude of $\sim 0.4$~mag. Modeling the observed amplitudes as a function of wavelength can provide a quantitative measure of the spot 
size and effective temperatures in the WTTSs \citep{bouvier1993}. However, 
a detailed analysis of their distribution on the stellar surface is not straight-forward due to the lack of geometrical constraints like line-of-sight inclination. 
Fig.~\ref{fig:tt_lcs} shows the $I$-band light curves of few TTSs in a specific time-range. Top panel shows a candidate WTT, V105, exhibiting near sinusoidal variations with 
similar peak values in each periodic cycle.  V139 shows periodic variation but variable peak brightness in different periodic cycles with MIR excess in CMD, and the amplitude 
variations are of the order of 1 mag in $I$. The quasi-periodic flux dips in the light curve of V139 could be driven by inner disk structures corotating with the star. It 
is classified as CTTS and the variability could also be due to dominant spots together with smaller extinction events. 
Another candidate CTTS, V184, displays a fading from the mean magnitude in October 2012 followed by a brightening phase, and achieves a state of high luminosity that seems to last 
over 50 days. It also exhibits several unresolved and possibly significant small-scale magnitude fluctuations during this brightest phase but the lowest luminosity state is recovered 
only in May 2013 after discontinuous observations. Similarly, V189 - a candidate CTTS, shows multiple extinction events of the order of 1~mag in $I$.

The amplitudes in $VI$-filters are shown in Fig.~\ref{fig:amp_col} for all variable sources. WTTSs have amplitudes smaller than $0.7$~mag while CTTSs exhibit large magnitude 
fluctuations up to $2.5$~mag. Four stars exhibit variability amplitudes significantly larger in the $I$-band than in the $V$-band. These cases exhibit lower-quality $V$-band 
light curves, which prevents an accurate determination of their luminosity maxima and minima. The colors of variability in all stars are also shown in the right panel of 
Fig.~\ref{fig:amp_col}. The variation in the luminosity and colors displays a range of slopes attributed to different cause of variability. Variables display redder colors 
at fainter epochs when the variability is driven by spot modulation. The variability amplitudes increase towards shorter wavelengths more steeply when the light 
curves are dominated by accretion spots than in the case of cold spot modulation \citep{vrba1993}. When the variability is driven by occultations due to opaque transiting material, little or 
no color variations are seen. The redder colors at fainter states are also observed in case of variability due to circumstellar extinction \citep{venuti2015}. 
In Fig.~\ref{fig:amp_col}, high-amplitude CTTSs display bluer colors than the smaller amplitude WTTSs. Most of the WTTSs stars show flatter slopes with little 
brightness variations while few CTTSs also occupy region towards high luminosity with smaller color variations. The MS/Field stars occupy a distinct region 
of the diagram with respect to the TTSs in our sample, and show very small variations both in their amplitude and colors.

\subsection{Bursters and faders}

In PMS stars, an increase in accretion rate from the circumstellar disk onto star can give rise to a significant burst in magnitude that lasts over hours to days \citep{cody2018}. Similarly,
short-duration brightening events with typical variation of a few tenths of a magnitude can also be seen that lasts typically up to a few hours. As in case of prototype AA Tau, repetitive 
fading of magnitudes for PMS stars occurred due to circumstellar extinction. \citet{findeisen2013} investigated bursters and faders with multiyear $R$-band time-series data from 
Palomar Transient Factory in North American and Pelican Nebulae. Six of those stars (2 bursters and 4 faders) are found common with our catalogue. V195 was a fader around mid-2011 in 
\citet{findeisen2013} while it shows a distinct brightening event lasting over 50 days at the end of 2012 in our photometry. Another fader, V182 shows a brightening event in November 
2012 while V180 shows multiple extinction events as large as $2$~mag. One burster V121, common with \citet{findeisen2013} also displays periodic variations in this work. 
While typical short-term, discrete bursts that occur within days are missed in our photometry, some PMS variables do show significant brightening within 10 days. For a simple 
quantitative measure, we define a burst if the brightness changes by more than 75 percentile of total amplitude within 10 days and occur in the epochs that are brighter than 
the median magnitude of the light curve. Only 12 CTTSs display these possible bursting events in the light curves. Individual fading or brightening/bursting events are commented in 
Table~\ref{tab:class} but we do not expect to observe any short-duration bursting events that only lasts 1 day, due to limited number of observations per night.

\subsection{Spectral energy distributions}
\begin{figure}
\centering
  \begin{tabular}{@{}c@{}}
    \includegraphics[width=.45\textwidth]{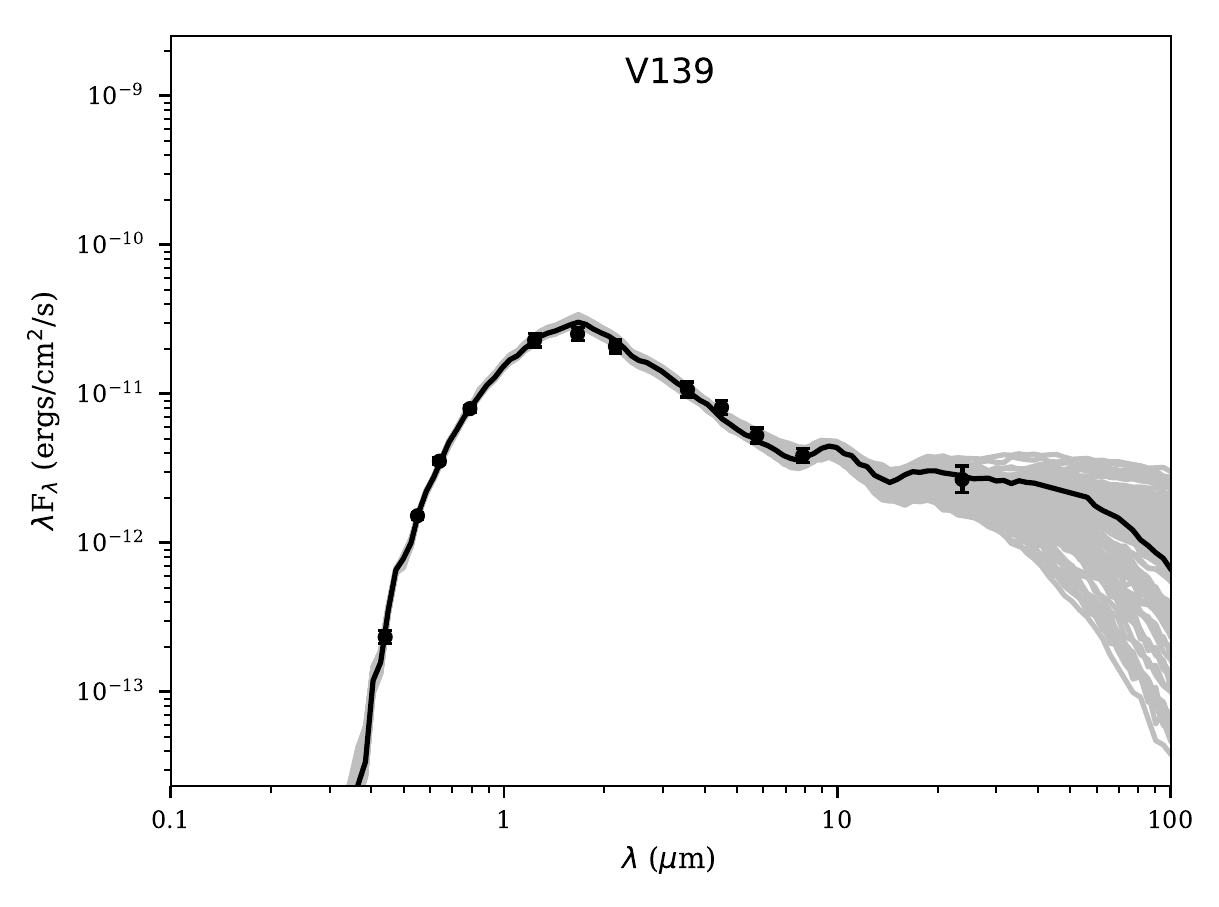} \\
    \includegraphics[width=.45\textwidth]{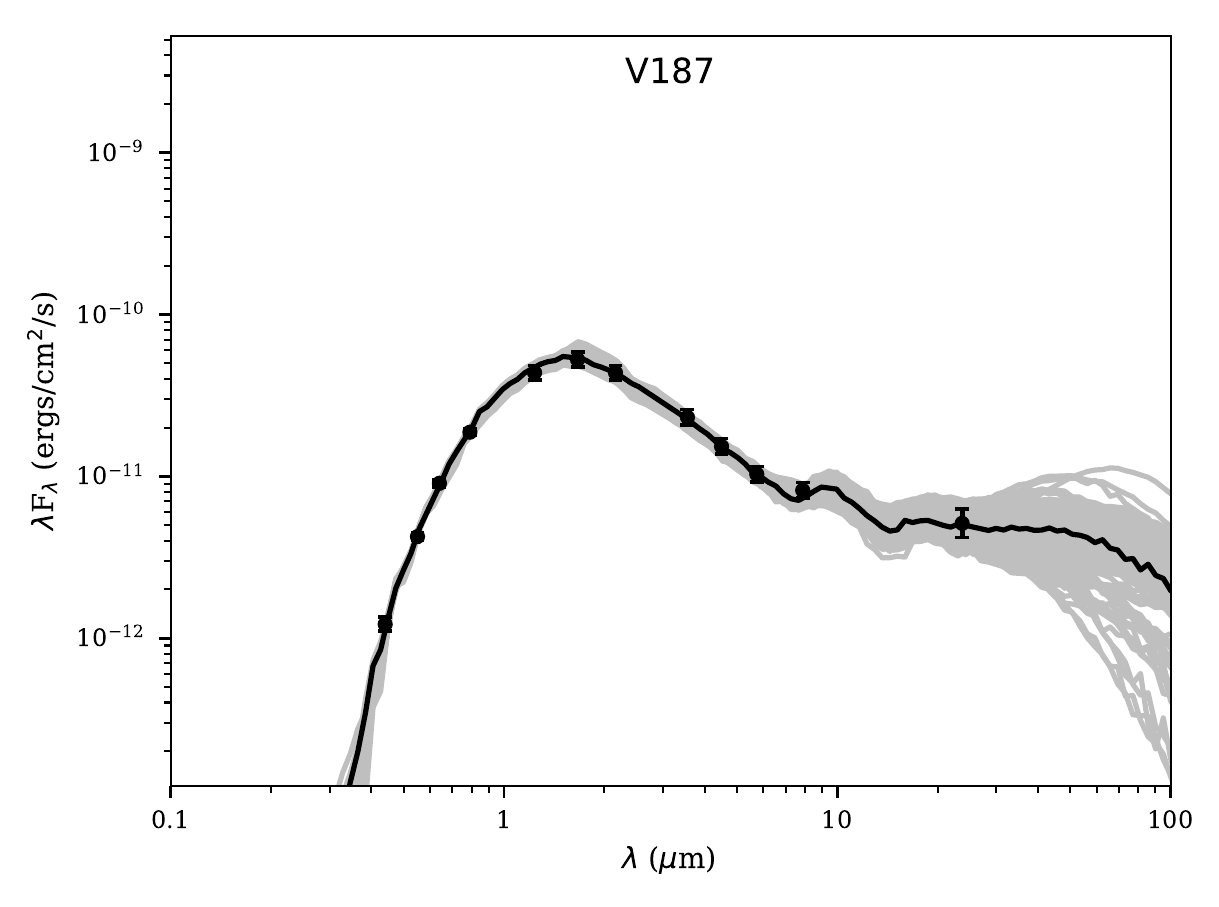} \\
  \end{tabular}
 \caption{The spectral energy distributions of two PMS variables. The black line shows the best-fit model while the grey lines display top 100 models that satisfy the
 criterion, $(\chi^2 - \chi^2_{best}) \le 2N$. The circles denote the observed fluxes at different wavelengths.}
 \label{fig:sed}
\end{figure}

The spectral energy distributions (SEDs) for the candidate PMS variables are constructed using the multiband photometric data compiled from the literature as discussed in Section 2. 
Optical $BVRI$ mean-magnitudes and random-epoch infrared magnitudes are converted into the mili-Jansky fluxes to construct observed SEDs. To infer the physical properties of these 
PMS stars, models of \citet{robitaille2006, robitaille2007} are fitted to the observed SEDs. The photometric uncertainties in the single-epoch infrared magnitudes are typically 
underestimated, therefore, a conservative estimate of uncertainties in the fluxes is adopted by adding 10$\%$ error in the quadrature to the flux errors derived from the 
uncertainties in infrared magnitudes. Since the amplitudes in infrared are typically smaller, these adopted uncertainties are reasonable to account for the magnitude variation within a random-epoch.
We emphasize that the SEDs generated from the radiative transfer models are subject to degeneracy as different combinations of parameters may
result in a similar fit to the observed SEDs \citep{robitaille2007}. This degeneracy could be remedied with spatially resolved observations at a range of wavelengths. 

In order to fit SED models to the observed fluxes, the mean distance to the cluster is adopted for all PMS variables. 
The extinction is allowed to vary from zero reddening to a maximum $A_V$ of 10~mag for SED-fitting. The range of $A_V$ is derived 
from the near-infrared color-color diagrams as discussed previously.  An upper limit of 
24 $\mu$m is imposed while fitting SEDs, although 70 $\mu$m flux is also available for a small sample of stars. The physical parameters of PMS stars can not be determined precisely if the 
number and range of wavelengths is small \citep{robitaille2007}. However, SED fits can still be used to constrain certain physical parameters depending on the availability of 
photometric data and a reasonable range of physical parameters can be interpreted. Therefore, SED models are fitted only if there are at least 10 flux measurements for a given PMS star and
select only those models for which, $(\chi^2 - \chi^2_{best}) \le 2N$, where $N$ is the number of data-points. Fig.~\ref{fig:sed} shows the SED fits of two variables in our sample. 
To estimate the physical parameters of the PMS stars, the mean values are estimated by weighted $e^{(-\chi^2/2)}$ of 100 best-fit models that satisfy the $\chi^2$ 
condition, and the standard deviation is adopted as their associated uncertainties. 

The typical ranges of mass and age estimates from the SED-fitting are consistent with those derived from the isochrone-fitting but the individual estimates may differ between the
two approaches. A median value of mass and age from SED-fitting is found to be $\sim 1.1 M_\odot$ and $\sim 2$~Myr, typically higher than isochrone based 
estimates. No mass-dependent trend is seen in the age estimates but the isochrone based mass and age estimates could be systematically smaller for low-mass stars 
\citep{hillenbrand2008, herczeg2015, pecaut2016}. The physical parameters obtained from the SED fits are listed in Table~\ref{tab:sed_phy} for the PMS stars for a relative comparison. 
Further, the luminosity and temperature estimates from the SED-fitting are compared with the values provided in the {\it Gaia} catalogue. While the luminosity for common 
stars correlate well, the temperatures are consistent within their uncertainties only in the 3000-5000K range. We also investigate possible correlations of variability 
periods and amplitudes with the physical parameters for PMS stars. No significant correlation is seen between the periods or amplitudes and the mass and age estimates 
for PMS stars. \citet{venuti2015} observed an anticorrelation between observed variability amplitudes in $u$ and $r$-band with stellar mass for WTTSs suggesting a 
uniform distribution of spots in more massive PMS stars. However, the limited sample of WTTSs in our sample preclude us from a detailed statistical analysis.

\begin{table}
\caption{Physical parameters of PMS stars based on SED fitting.}
\centering
\scalebox{0.9}{
\begin{tabular}{c r r r r}
\hline
ID & Mass              & Age  		&	Luminosity		&	Temperature\\
   & (M$_\odot$)       & (Myr.)  	& 	$\log(L/L_\odot)$		&	($\times 10^3$ K)\\
\hline
V102& 1.35$\pm$0.48 & 2.54$\pm$1.30 & 0.28$\pm$0.15 & 4.53$\pm$0.39 \\
V104& 1.72$\pm$0.36 & 4.44$\pm$3.07 & 0.76$\pm$0.16 & 5.25$\pm$0.69 \\
V105& 0.76$\pm$0.07 & 6.26$\pm$0.94 & -0.47$\pm$0.04 & 4.02$\pm$0.06 \\
V106& 1.92$\pm$0.41 & 4.17$\pm$2.71 & 0.76$\pm$0.17 & 5.38$\pm$0.67 \\
V107& 1.58$\pm$0.75 & 4.89$\pm$2.65 & 0.63$\pm$0.48 & 5.09$\pm$2.49 \\
V109& 1.59$\pm$0.56 & 3.94$\pm$3.39 & 0.57$\pm$0.29 & 4.83$\pm$0.62 \\
V112& 0.24$\pm$0.04 & 0.46$\pm$0.23 & -0.29$\pm$0.06 & 3.21$\pm$0.11 \\
V118& 0.47$\pm$0.26 & 0.51$\pm$2.01 & 0.33$\pm$0.49 & 3.67$\pm$0.32 \\
V120& 0.30$\pm$0.09 & 0.85$\pm$0.39 & -0.26$\pm$0.08 & 3.34$\pm$0.20 \\
V121& 2.00$\pm$0.48 & 2.84$\pm$2.62 & 1.05$\pm$0.17 & 5.35$\pm$0.94 \\
V122& 0.67$\pm$0.98 & 0.13$\pm$1.97 & 0.68$\pm$0.40 & 3.89$\pm$0.67 \\
V125& 0.20$\pm$0.03 & 1.35$\pm$0.48 & -0.60$\pm$0.04 & 3.11$\pm$0.07 \\
V126& 0.40$\pm$0.03 & 1.34$\pm$0.19 & -0.30$\pm$0.08 & 3.58$\pm$0.06 \\
V127& 0.42$\pm$0.91 & 0.37$\pm$5.72 & 0.23$\pm$0.25 & 3.64$\pm$1.22 \\
V128& 1.98$\pm$0.81 & 3.30$\pm$2.69 & 0.56$\pm$0.68 & 5.00$\pm$3.20 \\
V131& 0.45$\pm$0.06 & 1.62$\pm$0.17 & -0.34$\pm$0.06 & 3.68$\pm$0.07 \\
V135& 0.22$\pm$0.05 & 0.54$\pm$0.29 & -0.39$\pm$0.05 & 3.14$\pm$0.11 \\
V137& 0.32$\pm$0.05 & 1.27$\pm$0.11 & -0.33$\pm$0.06 & 3.41$\pm$0.11 \\
V138& 0.97$\pm$0.43 & 2.14$\pm$1.77 & 0.22$\pm$0.18 & 4.31$\pm$0.49 \\
V139& 1.60$\pm$0.30 & 2.76$\pm$1.18 & 0.42$\pm$0.10 & 4.72$\pm$0.27 \\
V143& 1.23$\pm$0.84 & 0.41$\pm$2.33 & 0.76$\pm$0.37 & 4.26$\pm$0.64 \\
V144& 0.90$\pm$0.68 & 0.82$\pm$1.05 & 0.45$\pm$0.24 & 4.12$\pm$0.47 \\
V146& 1.31$\pm$0.60 & 1.96$\pm$1.69 & 0.42$\pm$0.19 & 4.48$\pm$0.51 \\
V148& 1.04$\pm$0.53 & 1.99$\pm$1.64 & 0.26$\pm$0.33 & 4.29$\pm$0.36 \\
V149& 1.27$\pm$0.35 & 6.63$\pm$3.50 & 0.15$\pm$0.21 & 4.73$\pm$0.45 \\
V161& 1.12$\pm$1.00 & 0.27$\pm$1.92 & 0.90$\pm$0.28 & 4.19$\pm$1.22 \\
V162& 2.39$\pm$0.90 & 0.32$\pm$0.40 & 1.29$\pm$0.17 & 4.60$\pm$0.47 \\
V164& 0.46$\pm$0.71 & 0.68$\pm$3.21 & 0.14$\pm$0.27 & 3.71$\pm$0.78 \\
V167& 0.79$\pm$0.73 & 2.78$\pm$2.74 & 0.22$\pm$0.27 & 4.25$\pm$0.80 \\
V168& 1.60$\pm$0.75 & 2.18$\pm$2.12 & 0.87$\pm$0.24 & 4.89$\pm$0.86 \\
V169& 0.52$\pm$0.90 & 0.32$\pm$0.89 & 0.44$\pm$0.23 & 3.80$\pm$0.65 \\
V171& 1.12$\pm$0.64 & 0.62$\pm$0.49 & 0.66$\pm$0.16 & 4.23$\pm$0.42 \\
V178& 1.86$\pm$0.63 & 4.85$\pm$3.29 & 1.09$\pm$0.40 & 7.22$\pm$2.75 \\
V180& 2.92$\pm$1.50 & 1.07$\pm$2.32 & 1.11$\pm$0.48 & 5.02$\pm$1.00 \\
V182& 1.56$\pm$0.52 & 2.42$\pm$2.04 & 0.53$\pm$0.17 & 4.70$\pm$0.45 \\
V183& 2.08$\pm$0.82 & 4.84$\pm$1.92 & 1.12$\pm$0.54 & 5.72$\pm$3.89 \\
V184& 2.57$\pm$0.61 & 4.47$\pm$2.01 & 1.60$\pm$0.34 & 9.33$\pm$2.90 \\
V185& 0.73$\pm$0.57 & 2.59$\pm$3.28 & 0.04$\pm$0.30 & 3.94$\pm$0.82 \\
V186& 1.18$\pm$0.40 & 2.05$\pm$1.14 & 0.24$\pm$0.17 & 4.38$\pm$0.28 \\
V187& 1.76$\pm$0.45 & 2.37$\pm$2.40 & 0.70$\pm$0.17 & 4.86$\pm$0.57 \\
V189& 1.72$\pm$0.87 & 3.66$\pm$3.40 & 0.86$\pm$0.57 & 5.26$\pm$3.47 \\
V190& 2.64$\pm$0.47 & 4.89$\pm$1.98 & 1.63$\pm$0.24 & 9.57$\pm$2.34 \\
V192& 1.75$\pm$0.65 & 2.85$\pm$2.28 & 0.64$\pm$0.32 & 4.80$\pm$0.54 \\
V193& 2.60$\pm$0.89 & 3.51$\pm$1.75 & 1.69$\pm$0.44 & 8.04$\pm$3.39 \\
V194& 0.70$\pm$0.27 & 0.29$\pm$0.15 & 0.57$\pm$0.11 & 3.96$\pm$0.22 \\
\hline
\end{tabular}}
\label{tab:sed_phy}
\end{table}

\begin{figure}
  \begin{tabular}{@{}c@{}}
    \includegraphics[width=.5\textwidth]{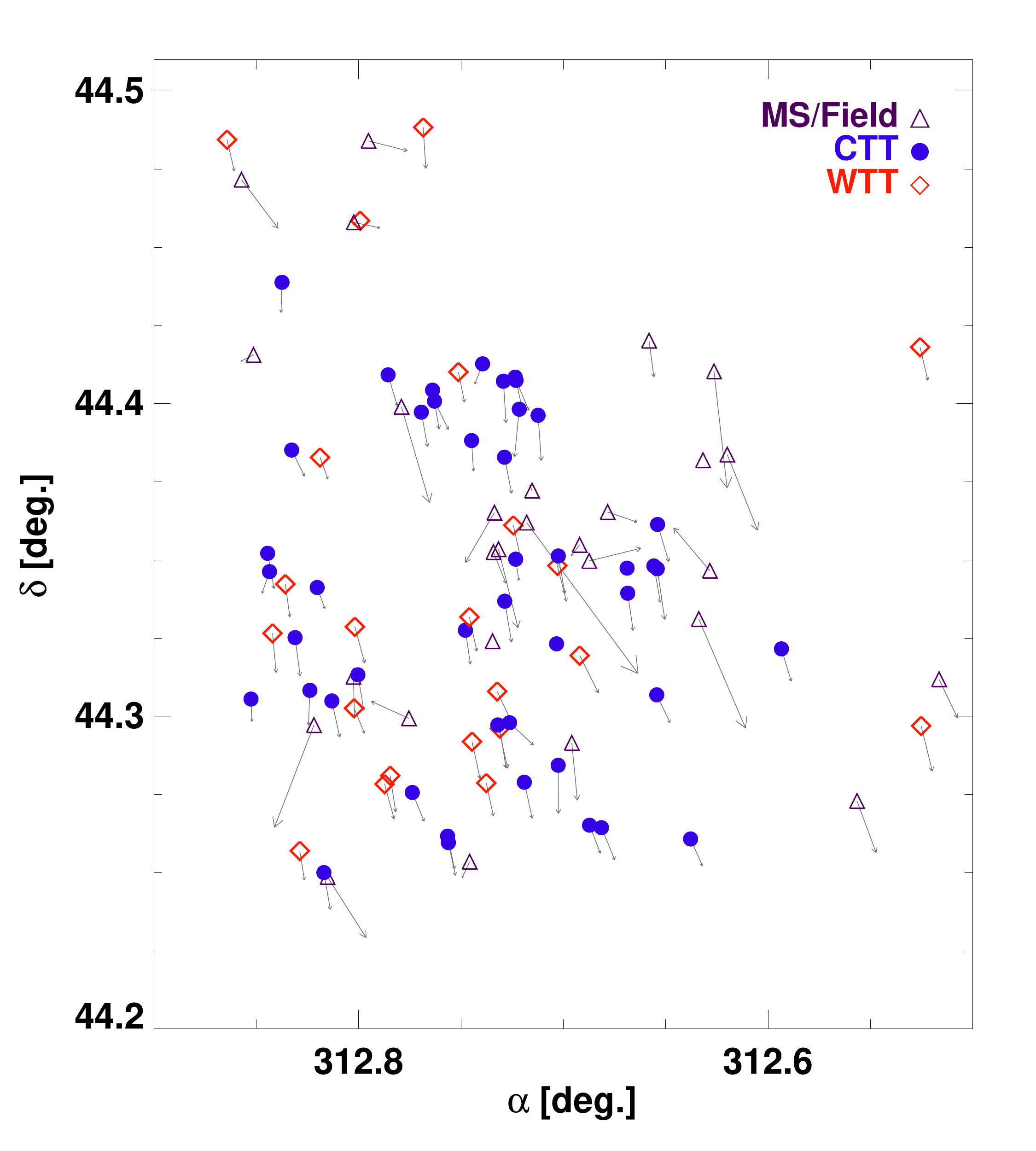} \\
  \end{tabular}
\caption{The spatial distribution for different class of variables along with their position 10$^4$ years ago pointed by the grey arrow.}
\label{fig:disc}
\end{figure}

\section{Discussion and Conclusions}
\label{sec:summary}

We presented a catalogue of optical time-series photometry of young stellar objects in a star-forming region Pelican Nebula (IC 5070). Our data provide a significant increase in the
number of pre-main-sequence variables in this region with multiband year-long optical photometry. Out of 95 variables in the targeted region, 67 objects are pre-main-sequence stars
classified based on the multiband color-magnitude and color-color diagrams. The five-parameter solutions from the recent {\it Gaia} data release are used to confirm the association
of variables with IC 5070 region using accurate proper motions and parallaxes. While the optical data are limited to at-most 3 epochs per night, a total of more than 250 epochs in 
each $VRI$-band allow us to further identify WTT and CTT candidates based on their light curve structure. Nearly $70\%$ of PMS stars display photometric variations similar to 
CTTSs and $30\%$ display strong periodic variations similar to WTTSs. Several CTTSs show significant extinction events and a few also exhibit the periodic variations. The amplitude 
variations for WTTSs are smaller than $0.4$~mag whereas the average amplitude variations for CTTSs are of 1~mag in $V$-band. CTTSs also display several fading and brightening 
events as large as $2.5$~mag in $I$-band.. The catalogue includes probable long-period variables displaying long-lasting ($>50$ days) brightening events. CTTSs display typical magnitude 
fluctuations up to three times the maximum variation seen in WTTSs in our sample.

Fig.~\ref{fig:disc} shows the spatial distribution of candidate T Tauri stars. All variables are distributed throughout the targeted field and no obvious clustering 
is seen for T Tauri stars. Most of the PMS variables have sub-solar masses and a median age of 2 Myr based on the isochrone-fitting to optical color-magnitude diagrams. 
The individual mass and age estimates may differ between the isochrone-based estimates and SED-fitting tools but the typical ranges of these physical parameters are 
consistent between the two approaches. We do not find any evidence of a correlation between amplitude and physical parameters. While this work provided a catalogue of 
variable sources, a detailed investigation into all pre-main sequence population and individual variable objects may be the subject of a future study. 

Variability studies of pre-main-sequence stars at shorter wavelengths can provide an insight into different physical properties of accreting and non-accreting young stellar objects. 
The candidate accreting CTTSs in our analysis display significantly higher variability than the disk-free WTTSs. The combined variations in the 
luminosity and colors for pre-main-sequence stars can be traced back to the root cause of their different variability signatures -  accretion, circumstellar extinction or spot-modulations.
Short-cadence time-series photometry and spectroscopic follow-up is desired to confirm the classification of T Tauri variables that would in turn allow a more detailed investigation into 
the root cause of variability for the pre-main-sequence stars presented in this analysis.

\section*{Acknowledgements}
We thank the anonymous referee for detailed and useful comments that have improved the quality of the paper.
AB acknowledges the research grant $\#11850410434$ awarded by the National Natural Science Foundation of China through a Research Fund for International Young Scientists,
and China Post-doctoral General Grant. NP acknowledges the financial support from the Department of Science and Technology, INDIA, through INSPIRE faculty award DST/IFA12/PH-36. 
HPS thanks Council of Scientific \& Industrial Research, India for grant $\#03$(1428)/18/EMR-II. We also thank Dr. Chow-Choong Ngeow for providing standard transformation equations, 
and Dr. Tapas Baug for sharing his calibration and plotting routines.
This research was support by the Munich Institute for Astro- and 
Particle Physics (MIAPP) of the DFG cluster of excellence ``Origin and Structure of the Universe''. This publication makes use of data from the Two Micron All Sky Survey 
(a joint project of the University of Massachusetts and the Infrared Processing and Analysis Center/California Institute of Technology, funded by the National Aeronautics 
and Space Administration and the National Science Foundation), and archival data obtained with the {\it Spitzer Space Telescope} and {\it Wide Infrared Survey Explorer} 
(operated by the Jet Propulsion Laboratory, California Institute of Technology, under contract with the NASA.

\bibliographystyle{aa}
\bibliography{/home/anupam/work/manuscripts/mybib_final}

\begin{thebibliography}{71}
\expandafter\ifx\csname natexlab\endcsname\relax\def\natexlab#1{#1}\fi

\bibitem[{{Alencar} {et~al.}(2010){Alencar}, {Teixeira}, {Guimar{\~a}es},
  {McGinnis}, {Gameiro}, {Bouvier}, {Aigrain}, {Flaccomio}, \&
  {Favata}}]{alencar2010}
{Alencar}, S.~H.~P., {Teixeira}, P.~S., {Guimar{\~a}es}, M.~M., {et~al.} 2010,
  A\&A, 519, A88

\bibitem[{{Ansdell} {et~al.}(2016){Ansdell}, {Gaidos}, {Rappaport}, {Jacobs},
  {LaCourse}, {Jek}, {Mann}, {Wyatt}, {Kennedy}, {Williams}, \&
  {Boyajian}}]{ansdell2016}
{Ansdell}, M., {Gaidos}, E., {Rappaport}, S.~A., {et~al.} 2016, ApJ, 816, 69

\bibitem[{{Bailer-Jones} {et~al.}(2018){Bailer-Jones}, {Rybizki}, {Fouesneau},
  {Mantelet}, \& {Andrae}}]{coryn2018}
{Bailer-Jones}, C.~A.~L., {Rybizki}, J., {Fouesneau}, M., {Mantelet}, G., \&
  {Andrae}, R. 2018, AJ, 156, 58

\bibitem[{{Bally} {et~al.}(2014){Bally}, {Ginsburg}, {Probst}, {Reipurth},
  {Shirley}, \& {Stringfellow}}]{bally2014}
{Bally}, J., {Ginsburg}, A., {Probst}, R., {et~al.} 2014, AJ, 148, 120

\bibitem[{{Baraffe} {et~al.}(2009){Baraffe}, {Chabrier}, \&
  {Gallardo}}]{baraffe2009}
{Baraffe}, I., {Chabrier}, G., \& {Gallardo}, J. 2009, ApJL, 702, L27

\bibitem[{{Baraffe} {et~al.}(2012){Baraffe}, {Vorobyov}, \&
  {Chabrier}}]{baraffe2012}
{Baraffe}, I., {Vorobyov}, E., \& {Chabrier}, G. 2012, ApJ, 756, 118

\bibitem[{{Bertout}(1989)}]{bertout1989}
{Bertout}, C. 1989, ARA\&A, 27, 351

\bibitem[{{Bessell} \& {Brett}(1988)}]{bessell1988}
{Bessell}, M.~S. \& {Brett}, J.~M. 1988, \pasp, 100, 1134

\bibitem[{{Bouvier} {et~al.}(1993){Bouvier}, {Cabrit}, {Fernandez}, {Martin},
  \& {Matthews}}]{bouvier1993}
{Bouvier}, J., {Cabrit}, S., {Fernandez}, M., {Martin}, E.~L., \& {Matthews},
  J.~M. 1993, A\&A, 272, 176

\bibitem[{{Bouvier} {et~al.}(1997){Bouvier}, {Forestini}, \&
  {Allain}}]{bouvier1997}
{Bouvier}, J., {Forestini}, M., \& {Allain}, S. 1997, A\&A, 326, 1023

\bibitem[{{Bressan} {et~al.}(2012){Bressan}, {Marigo}, {Girardi}, {Salasnich},
  {Dal Cero}, {Rubele}, \& {Nanni}}]{bressan2012}
{Bressan}, A., {Marigo}, P., {Girardi}, L., {et~al.} 2012, MNRAS, 427, 127

\bibitem[{{Cambr{\'e}sy} {et~al.}(2002){Cambr{\'e}sy}, {Beichman}, {Jarrett},
  \& {Cutri}}]{cambresy2002}
{Cambr{\'e}sy}, L., {Beichman}, C.~A., {Jarrett}, T.~H., \& {Cutri}, R.~M.
  2002, AJ, 123, 2559

\bibitem[{{Cody} \& {Hillenbrand}(2018)}]{cody2018}
{Cody}, A.~M. \& {Hillenbrand}, L.~A. 2018, AJ, 156, 71

\bibitem[{{Cody} {et~al.}(2014){Cody}, {Stauffer}, {Baglin}, {Micela},
  {Rebull}, {Flaccomio}, {Morales-Calder{\'o}n}, {Aigrain}, {Bouvier},
  {Hillenbrand}, {Gutermuth}, {Song}, {Turner}, {Alencar}, {Zwintz},
  {Plavchan}, {Carpenter}, {Findeisen}, {Carey}, {Terebey}, {Hartmann},
  {Calvet}, {Teixeira}, {Vrba}, {Wolk}, {Covey}, {Poppenhaeger}, {G{\"u}nther},
  {Forbrich}, {Whitney}, {Affer}, {Herbst}, {Hora}, {Barrado}, {Holtzman},
  {Marchis}, {Wood}, {Medeiros Guimar{\~a}es}, {Lillo Box}, {Gillen},
  {McQuillan}, {Espaillat}, {Allen}, {D'Alessio}, \& {Favata}}]{cody2014}
{Cody}, A.~M., {Stauffer}, J., {Baglin}, A., {et~al.} 2014, AJ, 147, 82

\bibitem[{{Cohen} {et~al.}(1981){Cohen}, {Frogel}, {Persson}, \&
  {Elias}}]{cohen1981}
{Cohen}, J.~G., {Frogel}, J.~A., {Persson}, S.~E., \& {Elias}, J.~H. 1981,
  \apj, 249, 481

\bibitem[{{Cutri} {et~al.}(2003){Cutri}, {Skrutskie}, {van Dyk}, {Beichman},
  {Carpenter}, {Chester}, {Cambresy}, {Evans}, {Fowler}, {Gizis}, {Howard},
  {Huchra}, {Jarrett}, {Kopan}, {Kirkpatrick}, {Light}, {Marsh}, {McCallon},
  {Schneider}, {Stiening}, {Sykes}, {Weinberg}, {Wheaton}, {Wheelock}, \&
  {Zacarias}}]{cutri2003}
{Cutri}, R.~M., {Skrutskie}, M.~F., {van Dyk}, S., {et~al.} 2003, {The IRSA
  2MASS All-Sky Point Source Catalog, NASA/IPAC Infrared Science Archive.}

\bibitem[{{David} {et~al.}(2019){David}, {Hillenbrand}, {Gillen}, {Cody},
  {Howell}, {Isaacson}, \& {Livingston}}]{david2019}
{David}, T.~J., {Hillenbrand}, L.~A., {Gillen}, E., {et~al.} 2019, ApJ, 872,
  161

\bibitem[{{Eyer} \& {Mowlavi}(2008)}]{eyer2008}
{Eyer}, L. \& {Mowlavi}, N. 2008, in Journal of Physics Conference Series, Vol.
  118, Journal of Physics Conference Series, 012010

\bibitem[{{Fernandes} {et~al.}(2018){Fernandes}, {Long}, {Pikhartova}, {Sitko},
  {Grady}, {Russell}, {Luria}, {Tyler}, {Bayyari}, {Danchi}, \&
  {Wisniewski}}]{fernandes2018}
{Fernandes}, R.~B., {Long}, Z.~C., {Pikhartova}, M., {et~al.} 2018, ApJ, 856,
  103

\bibitem[{{Findeisen} {et~al.}(2013){Findeisen}, {Hillenbrand}, {Ofek},
  {Levitan}, {Sesar}, {Laher}, \& {Surace}}]{findeisen2013}
{Findeisen}, K., {Hillenbrand}, L., {Ofek}, E., {et~al.} 2013, ApJ, 768, 93

\bibitem[{{Froebrich} {et~al.}(2018){Froebrich}, {Scholz}, {Campbell-White},
  {Crumpton}, {DArcy}, {Makin}, {Zegmott}, {Billington}, {Hibbert}, {Newport},
  \& {Fisher}}]{froebrich2018}
{Froebrich}, D., {Scholz}, A., {Campbell-White}, J., {et~al.} 2018, Research
  Notes of the AAS, 2, 61

\bibitem[{{Gaia Collaboration}(2018)}]{gaia2018cat}
{Gaia Collaboration}. 2018, VizieR Online Data Catalog, 1345

\bibitem[{{Gaia Collaboration} {et~al.}(2018){Gaia Collaboration}, {Brown},
  {Vallenari}, {Prusti}, {de Bruijne}, {Babusiaux}, \&
  {Bailer-Jones}}]{gaia2018}
{Gaia Collaboration}, {Brown}, A.~G.~A., {Vallenari}, A., {et~al.} 2018, ArXiv
  e-prints, 1804.09365

\bibitem[{{Gillen} {et~al.}(2017){Gillen}, {Aigrain}, {Terquem}, {Bouvier},
  {Alencar}, {Gandolfi}, {Stauffer}, {Cody}, {Venuti}, {Almeida}, {Micela},
  {Favata}, \& {Deeg}}]{gillen2017}
{Gillen}, E., {Aigrain}, S., {Terquem}, C., {et~al.} 2017, A\&A, 599, A27

\bibitem[{{Girardi} {et~al.}(2002){Girardi}, {Bertelli}, {Bressan}, {Chiosi},
  {Groenewegen}, {Marigo}, {Salasnich}, \& {Weiss}}]{girardi2002}
{Girardi}, L., {Bertelli}, G., {Bressan}, A., {et~al.} 2002, A\&A, 391, 195

\bibitem[{{Grankin} {et~al.}(2008){Grankin}, {Bouvier}, {Herbst}, \&
  {Melnikov}}]{grankin2008}
{Grankin}, K.~N., {Bouvier}, J., {Herbst}, W., \& {Melnikov}, S.~Y. 2008, A\&A,
  479, 827

\bibitem[{{Guieu} {et~al.}(2009){Guieu}, {Rebull}, {Stauffer}, {Hillenbrand},
  {Carpenter}, {Noriega-Crespo}, {Padgett}, {Cole}, {Carey}, {Stapelfeldt}, \&
  {Strom}}]{gui2009}
{Guieu}, S., {Rebull}, L.~M., {Stauffer}, J.~R., {et~al.} 2009, ApJ, 697, 787

\bibitem[{{Guo} {et~al.}(2018){Guo}, {Herczeg}, {Jose}, {Fu}, {Chiang},
  {Grankin}, {Michel}, {Kesh Yadav}, {Liu}, {Chen}, {Li}, {Xue}, {Niu},
  {Subramaniam}, {Sharma}, {Prasert}, {Flores-Fajardo}, {Castro}, \&
  {Altamirano}}]{guo2018}
{Guo}, Z., {Herczeg}, G.~J., {Jose}, J., {et~al.} 2018, ApJ, 852, 56

\bibitem[{{Henden} {et~al.}(2016){Henden}, {Templeton}, {Terrell}, {Smith},
  {Levine}, \& {Welch}}]{henden2016}
{Henden}, A.~A., {Templeton}, M., {Terrell}, D., {et~al.} 2016, VizieR Online
  Data Catalog, 2336

\bibitem[{{Herbig}(1962)}]{herbig1962}
{Herbig}, G.~H. 1962, Advances in Astronomy and Astrophysics, 1, 47

\bibitem[{{Herbig}(1977)}]{herbig1977}
{Herbig}, G.~H. 1977, ApJ, 214, 747

\bibitem[{{Herbst} {et~al.}(2007){Herbst}, {Eisloffel}, {Mundt}, \&
  {Scholz}}]{herbst2007}
{Herbst}, W., {Eisloffel}, J., {Mundt}, R., \& {Scholz}, A. 2007, in Protostars
  and Planets V, eds. B. Reipurth, D. Jewitt, K. Keil, University of Arizona
  Press, Tucson, 297

\bibitem[{{Herbst} {et~al.}(1994){Herbst}, {Herbst}, {Grossman}, \&
  {Weinstein}}]{herbst1994}
{Herbst}, W., {Herbst}, D.~K., {Grossman}, E.~J., \& {Weinstein}, D. 1994, AJ,
  108, 1906

\bibitem[{{Herczeg} \& {Hillenbrand}(2014)}]{herczeg2014}
{Herczeg}, G.~J. \& {Hillenbrand}, L.~A. 2014, ApJ, 786, 97

\bibitem[{{Herczeg} \& {Hillenbrand}(2015)}]{herczeg2015}
{Herczeg}, G.~J. \& {Hillenbrand}, L.~A. 2015, ApJ, 808, 23

\bibitem[{{Hillenbrand} {et~al.}(2008){Hillenbrand}, {Carpenter}, {Kim},
  {Meyer}, {Backman}, {Moro-Mart{\'{\i}}n}, {Hollenbach}, {Hines}, {Pascucci},
  \& {Bouwman}}]{hillenbrand2008}
{Hillenbrand}, L.~A., {Carpenter}, J.~M., {Kim}, J.~S., {et~al.} 2008, ApJ,
  677, 630

\bibitem[{{Ibryamov} {et~al.}(2018){Ibryamov}, {Semkov}, {Milanov}, \&
  {Peneva}}]{Ibryamov2018}
{Ibryamov}, S., {Semkov}, E., {Milanov}, T., \& {Peneva}, S. 2018, ArXiv
  e-prints [\eprint[arXiv]{1805.11745}]

\bibitem[{{Ikeda} {et~al.}(2008){Ikeda}, {Sugitani}, {Watanabe}, {Fukuda},
  {Tamura}, {Nakajima}, {Pickles}, {Nagashima}, {Nagayama}, {Nakaya}, {Nakano},
  \& {Nagata}}]{ikeda2008}
{Ikeda}, H., {Sugitani}, K., {Watanabe}, M., {et~al.} 2008, AJ, 135, 2323

\bibitem[{{Joy}(1945)}]{joy1945}
{Joy}, A.~H. 1945, ApJ, 102, 168

\bibitem[{{K{\'o}sp{\'a}l} {et~al.}(2011){K{\'o}sp{\'a}l}, {{\'A}brah{\'a}m},
  {Acosta-Pulido}, {Ar{\'e}valo Morales}, {Carnerero}, {Elek}, {Kelemen},
  {Kun}, {P{\'a}l}, {Szak{\'a}ts}, \& {Vida}}]{kospal2011}
{K{\'o}sp{\'a}l}, {\'A}., {{\'A}brah{\'a}m}, P., {Acosta-Pulido}, J.~A.,
  {et~al.} 2011, A\&A, 527, A133

\bibitem[{{Landolt}(2009)}]{landolt2009}
{Landolt}, A.~U. 2009, AJ, 137, 4186

\bibitem[{{Lata} {et~al.}(2016){Lata}, {Pandey}, {Panwar}, {Chen}, {Samal}, \&
  {Pandey}}]{lata2016}
{Lata}, S., {Pandey}, A.~K., {Panwar}, N., {et~al.} 2016, MNRAS, 456, 2505

\bibitem[{{Laugalys} {et~al.}(2007){Laugalys}, {Strai{\v z}ys}, {Vrba}, {{\v
  C}ernis}, {Kazlauskas}, {Boyle}, \& {Philip}}]{laugalys2007}
{Laugalys}, V., {Strai{\v z}ys}, V., {Vrba}, F.~J., {et~al.} 2007, Baltic
  Astronomy, 16, 349

\bibitem[{{Lomb}(1976)}]{lomb1976}
{Lomb}, N.~R. 1976, APSS, 39, 447

\bibitem[{{Lucas} {et~al.}(2008){Lucas}, {Hoare}, {Longmore}, {Schr{\"o}der},
  {Davis}, {Adamson}, {Bandyopadhyay}, {de Grijs}, {Smith}, {Gosling},
  {Mitchison}, {G{\'a}sp{\'a}r}, {Coe}, {Tamura}, {Parker}, {Irwin}, {Hambly},
  {Bryant}, {Collins}, {Cross}, {Evans}, {Gonzalez-Solares}, {Hodgkin},
  {Lewis}, {Read}, {Riello}, {Sutorius}, {Lawrence}, {Drew}, {Dye}, \&
  {Thompson}}]{lucas2008}
{Lucas}, P.~W., {Hoare}, M.~G., {Longmore}, A., {et~al.} 2008, MNRAS, 391, 136

\bibitem[{{Luri} {et~al.}(2018){Luri}, {Brown}, {Sarro}, {Arenou},
  {Bailer-Jones}, {Castro-Ginard}, {de Bruijne}, {Prusti}, {Babusiaux}, \&
  {Delgado}}]{luri2018}
{Luri}, X., {Brown}, A.~G.~A., {Sarro}, L.~M., {et~al.} 2018, A\&A, 616, A9

\bibitem[{{Messina} {et~al.}(2017){Messina}, {Parihar}, \&
  {Distefano}}]{messina2017}
{Messina}, S., {Parihar}, P., \& {Distefano}, E. 2017, MNRAS, 468, 931

\bibitem[{{Meyer} {et~al.}(1997){Meyer}, {Calvet}, \&
  {Hillenbrand}}]{meyer1997}
{Meyer}, M.~R., {Calvet}, N., \& {Hillenbrand}, L.~A. 1997, AJ, 114, 288

\bibitem[{{Ogura} {et~al.}(2002){Ogura}, {Sugitani}, \& {Pickles}}]{ogura2002}
{Ogura}, K., {Sugitani}, K., \& {Pickles}, A. 2002, \aj, 123, 2597

\bibitem[{{Panwar} {et~al.}(2014){Panwar}, {Chen}, {Pandey}, {Samal}, {Ogura},
  {Ojha}, {Jose}, \& {Bhatt}}]{panwar2014}
{Panwar}, N., {Chen}, W.~P., {Pandey}, A.~K., {et~al.} 2014, MNRAS, 443, 1614

\bibitem[{{Pecaut} \& {Mamajek}(2016)}]{pecaut2016}
{Pecaut}, M.~J. \& {Mamajek}, E.~E. 2016, MNRAS, 461, 794

\bibitem[{{Poljan{\v c}i{\'c}} {et~al.}(2014){Poljan{\v c}i{\'c}}, {Jurdana-{\v
  S}epi{\'c}}, {Semkov}, {Ibryamov}, {Peneva}, \& {Tsvetkov}}]{poljan2014}
{Poljan{\v c}i{\'c}}, B.~I., {Jurdana-{\v S}epi{\'c}}, R., {Semkov}, E.~H.,
  {et~al.} 2014, A\&A, 568, A49

\bibitem[{{Rebull} {et~al.}(2011){Rebull}, {Guieu}, {Stauffer}, {Hillenbrand},
  {Noriega-Crespo}, {Stapelfeldt}, {Carey}, {Carpenter}, {Cole}, {Padgett},
  {Strom}, \& {Wolff}}]{rebull2011}
{Rebull}, L.~M., {Guieu}, S., {Stauffer}, J.~R., {et~al.} 2011, ApJs, 193, 25

\bibitem[{{Reipurth} \& {Schneider}(2008)}]{reipurth2008}
{Reipurth}, B. \& {Schneider}, N. 2008, {Star Formation and Young Clusters in
  Cygnus}, Vol. {Reipurth}, B., 36

\bibitem[{{Robitaille} {et~al.}(2007){Robitaille}, {Whitney}, {Indebetouw}, \&
  {Wood}}]{robitaille2007}
{Robitaille}, T.~P., {Whitney}, B.~A., {Indebetouw}, R., \& {Wood}, K. 2007,
  ApJs, 169, 328

\bibitem[{{Robitaille} {et~al.}(2006){Robitaille}, {Whitney}, {Indebetouw},
  {Wood}, \& {Denzmore}}]{robitaille2006}
{Robitaille}, T.~P., {Whitney}, B.~A., {Indebetouw}, R., {Wood}, K., \&
  {Denzmore}, P. 2006, ApJs, 167, 256

\bibitem[{{Rodriguez} {et~al.}(2017){Rodriguez}, {Ansdell}, {Oelkers},
  {Cargile}, {Gaidos}, {Cody}, {Stevens}, {Somers}, {James}, {Beatty},
  {Siverd}, {Lund}, {Kuhn}, {Gaudi}, {Pepper}, \& {Stassun}}]{rodriguez2017}
{Rodriguez}, J.~E., {Ansdell}, M., {Oelkers}, R.~J., {et~al.} 2017, ApJ, 848,
  97

\bibitem[{{Scargle}(1982)}]{scargle1982}
{Scargle}, J.~D. 1982, ApJ, 263, 835

\bibitem[{{Schwarzenberg-Czerny}(1989)}]{schwarz1989}
{Schwarzenberg-Czerny}, A. 1989, MNRAS, 241, 153

\bibitem[{{Semkov}(2011)}]{semkov2011}
{Semkov}, E.~H. 2011, Bulgarian Astronomical Journal, 15, 65

\bibitem[{{Siess} {et~al.}(2000){Siess}, {Dufour}, \& {Forestini}}]{siess2000}
{Siess}, L., {Dufour}, E., \& {Forestini}, M. 2000, A\&A, 358, 593

\bibitem[{{Soderblom} {et~al.}(2014){Soderblom}, {Hillenbrand}, {Jeffries},
  {Mamajek}, \& {Naylor}}]{soderblom2014}
{Soderblom}, D.~R., {Hillenbrand}, L.~A., {Jeffries}, R.~D., {Mamajek}, E.~E.,
  \& {Naylor}, T. 2014, Protostars and Planets VI, 219

\bibitem[{{Stassun} {et~al.}(2014){Stassun}, {Feiden}, \&
  {Torres}}]{stassun2014}
{Stassun}, K.~G., {Feiden}, G.~A., \& {Torres}, G. 2014, NewAR, 60, 1

\bibitem[{{Stellingwerf}(1978)}]{stellingwerf1978}
{Stellingwerf}, R.~F. 1978, ApJ, 224, 953

\bibitem[{{Stelzer}(2015)}]{stelzer2015}
{Stelzer}, B. 2015, Astronomische Nachrichten, 336, 493

\bibitem[{{Stetson}(1987)}]{stetson1987}
{Stetson}, P.~B. 1987, PASP, 99, 191

\bibitem[{{Stetson}(1993)}]{stetson1993}
{Stetson}, P.~B. 1993, in IAU Colloq. 136: Stellar Photometry - Current
  Techniques and Future Developments, ed. C.~J. {Butler} \& I.~{Elliott}, Vol.
  136, 291

\bibitem[{{Stetson}(1996)}]{stetson1996}
{Stetson}, P.~B. 1996, PASP, 108, 851

\bibitem[{{Venuti} {et~al.}(2015){Venuti}, {Bouvier}, {Irwin}, {Stauffer},
  {Hillenbrand}, {Rebull}, {Cody}, {Alencar}, {Micela}, {Flaccomio}, \&
  {Peres}}]{venuti2015}
{Venuti}, L., {Bouvier}, J., {Irwin}, J., {et~al.} 2015, A\&A, 581, A66

\bibitem[{{Vrba} {et~al.}(1993){Vrba}, {Chugainov}, {Weaver}, \&
  {Stauffer}}]{vrba1993}
{Vrba}, F.~J., {Chugainov}, P.~F., {Weaver}, W.~B., \& {Stauffer}, J.~S. 1993,
  AJ, 106, 1608

\bibitem[{{Zhang} {et~al.}(2014){Zhang}, {Xu}, \& {Yang}}]{zhang2014}
{Zhang}, S., {Xu}, Y., \& {Yang}, J. 2014, AJ, 147, 46

\end{thebibliography}


\section{Supplementary material}

\subsection{Multiband light-curves}
\begin{figure*}
\centering
\includegraphics[width=1.0\textwidth]{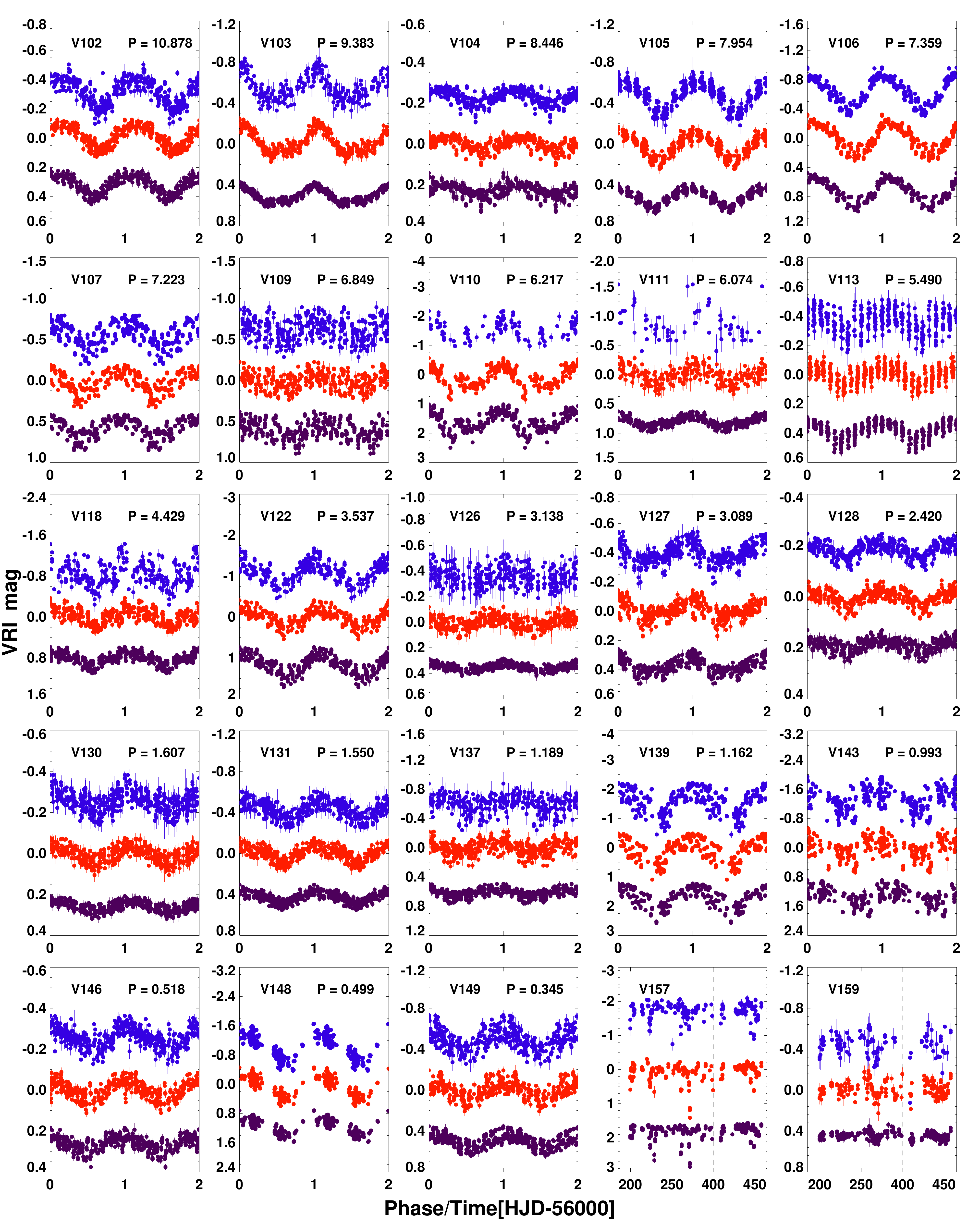}
\caption{Phase/time light curves of variable stars.}
\label{fig:lc0}
\end{figure*}

\begin{figure*}
\centering
\includegraphics[width=1.0\textwidth]{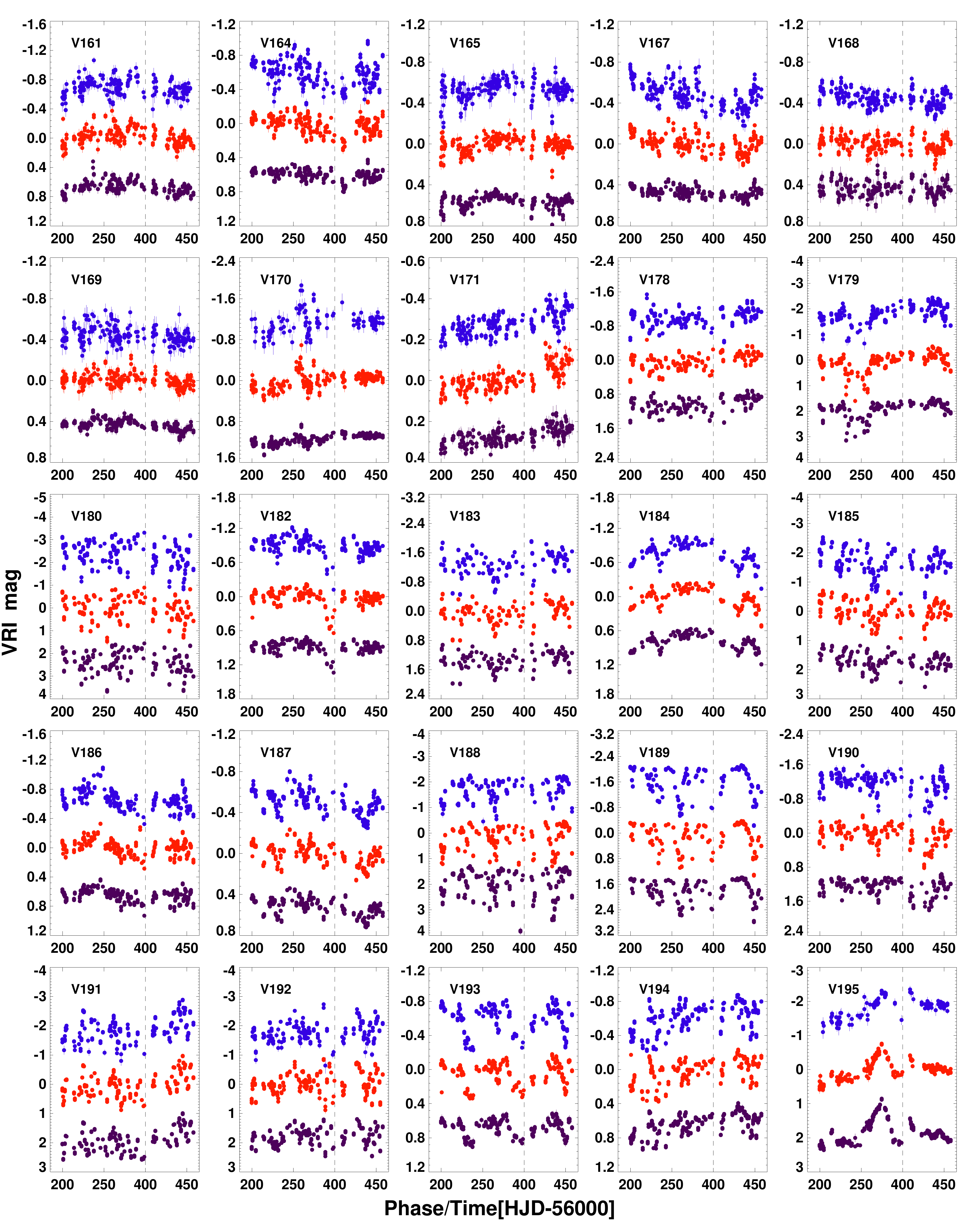}
\caption{Same as Fig.~\ref{fig:lc0}.}
\label{fig:lc1}
\end{figure*}

The light curves of pre-main-sequence variable stars observed in all three ($VRI$) bands are shown in Fig.~\ref{fig:lc0}-\ref{fig:lc1}. $R$-band light curves are shown in the middle (red) of each panel.
Light curves in $V/I$-bands are displayed above/below (blue/violet) the $R$-band, and are offset by $R$-band amplitude for visualization purposes. The star ID and periods are mentioned 
at the top of each panel. In case of time light curves, the $x$-axis is offset by $HJD-56000$, while the vertical dashed line separates the observations in two different seasons. 
All light curves are normalized with zero-mean for a relative comparison and plotting purposes. These figures are available online as supplementary material.

\end{document}